\documentstyle[prb,aps,multicol,eqsecnum,psfig]{revtex}

\begin{document}

\draft

\title{   Spectral and transport properties of doped Mott-Hubbard systems
                       with incommensurate magnetic order }

\author{  Marcus Fleck and Alexander I. Lichtenstein }
\address{ Max-Planck-Institut f\"{u}r Festk\"{o}rperforschung,
          Heisenbergstrasse 1, D-70569 Stuttgart, Federal Republic of Germany}

\author{  Andrzej M. Ole\'s }
\address{ Institute of Physics, Jagellonian University, Reymonta 4, 
          PL-30059 Krak\'ow, Poland, \\
          and Max-Planck-Institut f\"{u}r Festk\"{o}rperforschung,
          Heisenbergstrasse 1, D-70569 Stuttgart, Federal Republic of Germany}

\author{  Lars Hedin }
\address{ Department of Theoretical Physics, University of Lund,
          Solvegatan 14A, S-22362 Lund, Sweden \\
          and Max-Planck-Institut f\"{u}r Festk\"{o}rperforschung,
          Heisenbergstrasse 1, D-70569 Stuttgart, Federal Republic of Germany}

\date{November 11, 1998}
\date{\today}
\maketitle

\begin{abstract}
We present spectral and optical properties of the Hubbard model on a
two-dimensional square lattice using a generalization of dynamical 
mean-field theory to magnetic states in finite dimension. The self-energy 
includes the effect of spin fluctuations and screening of the Coulomb 
interaction due to particle-particle scattering. At half-filling the 
quasiparticles reduce the width of the Mott-Hubbard `gap' and have dispersions 
and spectral weights that agree remarkably well with quantum Monte Carlo and 
exact diagonalization calculations. Away from half-filling we consider 
incommensurate magnetic order with a varying local spin direction, and derive 
the photoemission and optical spectra. The incommensurate magnetic order leads 
to a pseudogap which opens at the Fermi energy and coexists with a large 
Mott-Hubbard gap. The quasiparticle states survive in the doped systems, but 
their dispersion is modified with the doping and a rigid band picture does 
not apply. Spectral weight in the optical conductivity is transferred to 
lower energies and the Drude weight increases linearly with increasing 
doping. We show that incommensurate magnetic order leads also to mid-gap 
states in the optical spectra and to decreased scattering rates in the 
transport processes, in qualitative agreement with the experimental 
observations in doped systems. The gradual disappearence of the spiral 
magnetic order and the vanishing pseudogap with increasing temperature is 
found to be responsible for the linear resistivity. We discuss the possible
reasons why these results may only partially explain the features observed 
in the optical spectra of high temperature superconductors.
\end{abstract}



\begin{multicols}{2} 

\section{Introduction}
\label{sec:intro}

In the past decade the interest in the physical properties of correlated
electronic systems has greatly increased. One reason is the strong local 
correlations on transition metal ions in cuprate superconductors and 
manganites, and the corresponding unusual properties of these compounds. 
The parent undoped compounds are Mott-Hubbard or charge transfer insulators,
while doping leads to correlated metals in which the kinetic energy of
charge carriers competes with magnetic order.\cite{Ima98} One of the most
spectacular consequences is the onset of high-temperature superconductivity
in the cuprates. It is believed that a satisfactory description of the
normal phase properties is a prerequisite for the understanding of the
microscopic mechanism of pairing in high-temperature superconductors. The
electronic states in CuO$_2$ planes of cuprate superconductors are usually
described in terms of the Emery model which includes the hybridization
between Cu($3d_{x^2-y^2}$) and O($2p_{x(y)}$) states.\cite{Var87} 
However, hole doping leads to the formation of local Zhang-Rice singlets,
\cite{Zha88} and the essential excitations in the cuprates within a window 
of a few eV around the chemical potential are well reproduced using the 
effective two-dimensional (2D) Hubbard model with extended hopping
\cite{Fei92} and a large local Coulomb interaction $U$, as shown by 
various numerical studies of the $t$-$J$ and Hubbard model.\cite{Dag94} 
 
Recently it was shown \cite{Fei92,Bal95,And95} that the effective Hubbard 
model has to include hopping beyond the nearest-neighbors. The second 
nearest-neighbor hopping changes the dispersion of the quasiparticle (QP)
states, and is therefore crucial for understanding angular resolved
photoemission (ARPES) data of the antiferromagnetic (AF) insulator
Sr$_2$CuO$_2$Cl$_2$.\cite{Bal95} Both second and third neighbor
hopping parameters follow from the down-folding procedure in electronic
structure calculations,\cite{And95} and influence the shape of the Fermi
surface. They have a particular relation to the value of the superconducting
transition temperature at optimal doping.\cite{Fei96}

The superconductivity occurs in the cuprates under doping $\delta=1-n$ of
a half-filled ($n=1$) AF insulator, and is accompanied by a gradual
modification of the magnetic order. The nature of magnetic correlations in
doped materials is therefore a central issue in the theory of the cuprate
superconductors. Undoped La$_2$CuO$_4$ is a {\it commensurate} AF insulator,
while doping by Sr into La$_{2-x}$Sr$_x$CuO$_4$ results in short-range AF
order within {\it incommensurate} magnetic structures.\cite{Tra97,Kas98}
Such an incommensurate magnetic order was indeed found
analytically,\cite{Shr89,Kub93} in Hartree-Fock (HF),\cite{Zaa89,Arr91} and
in a slave-boson approximation.\cite{Arr91,Kan90,Fre92,Iga92} However, in
order to understand the transport properties one has to go beyond an effective
single-particle description and include the dynamics due to local electron
correlations.

A sufficiently accurate treatment of local electron correlations remains one
of the challenging problems in modern solid state theory.
Although an important progress in the present understanding of strongly
correlated fermion systems occurred recently due to numerical methods, such
as quantum Monte Carlo (QMC) and exact diagonalization (ED), an analytic
treatment that maintains local correlations is needed to investigate the
consequences of strong correlations in the thermodynamic limit.
An attractive possibility is the limit of large spatial dimension
($d=\infty$), where the diagrams in the perturbative expansion collapse to
a single site and the fermion dynamics is described by a {\em local
self-energy\/}.\cite{Met89} This allows a mapping of lattice models onto 
quantum impurity models, which can then be solved self-consistently
using the dynamical mean-field theory (DMFT).\cite{Geo96}

The DMFT was quite successful for the Hubbard model with nearest-neighbor
hopping $t$ at half-filling, where it predicts the Mott transition to 
the insulating state ($n=1$).\cite{Jar92} This was also found by Logan, 
Eastwood, and Tusch \cite{Log96} for the $d=\infty$ case using an analytic 
method. Attempts to use DMFT at arbitrary filling, however, made it clear 
that the local self-energy becomes particularly important in systems with 
magnetic long-range order (LRO), which are easily destabilized when the 
correlation effects are overestimated. The self-energy therefore plays a 
decisive role and has to be described {\it beyond\/} second order 
perturbation theory (SOPT).\cite{Kaj96,Pot97} This has made the application 
of DMFT to magnetically ordered systems notoriously difficult. 
Recently we have shown that the screening of local Coulomb interaction by
the particle-particle diagrams plays a crucial role in stabilizing the
incommensurate magnetic LRO in doped systems.\cite{Fle98}

The advantage of using DMFT becomes clear by looking at the single hole 
problem, which can be solved exactly in the $d\to\infty$ limit.\cite{Str92} 
The method becomes exact because the quantum fluctuations are of higher 
order in the $1/d$-expansion than the leading potential term which 
originates from the Ising part of the superexchange interaction $J=4t^2/U$. 
Therefore, applying DMFT to the $d=2$ case might still capture the essential
features that result from the coupling of a moving hole to local spin
fluctuations. We will show below that in fact such quantities as the spectral
function, the QP band, and the size of the QP spectral weight are well
reproduced within the DMFT, which for a single hole includes only those 
processes which are present in the $t-J_z$ model. Although this approach
becomes exact only in the $d\to\infty$ limit,\cite{Str92} it gives a 
sufficient accuracy of the one-particle spectral function even in finite 
dimension $d=2$.\cite{Pru96} The DMFT allows us to calculate the optical 
conductivity in 
the $d=\infty$ limit of Metzner and Vollhardt \cite{Met89} from the knowledge 
of the local self-energy without further approximations.\cite{Khu90} The 
studies performed in this limit for the nonmagnetic systems already allowed 
a qualitative reproduction of such experimental observations in the cuprates 
as the increase of the Drude peak with doping, and a temperature and doping 
dependent mid-infrared peak.\cite{Jar95,Pru95} 

The paper is organized as follows. The self-consistent procedure to determine
a local self-energy within the DMFT is introduced in Sec. II. It consists of
the HF potential and the dynamical part due to spin fluctuations which uses
a Coulomb interaction renormalized by particle-particle scattering. The 
formalism to calculate the one-particle and optical excitation spectra in the 
spin spiral (SS) states is developed in Sec. III. Next we analyze the 
numerical results for the one-particle spectral properties in Sec. IV, where 
we show how they change with doping and with increasing temperature. 
The optical properties are presented in Sec. V; there we discuss the new 
effects in the optical conductivity, scattering rate and the effective mass 
which arise due to extended hopping and by increasing the value of Coulomb 
interaction $U$. Sec. VI presents a short summary and conclusions.

\section{Dynamical mean-field theory for spin spiral order}
\label{sec:theory}

\subsection{Dynamical mean-field equations}
\label{sec:dmft}

We consider the spectral and optical properties of the minimal model for
strongly correlated electrons in high-temperature superconductors, the 
Hubbard model with extended hopping,\cite{Fei92,And95}
\begin{equation}
H = -\sum_{ij\sigma}t_{ij}a^{\dagger}_{i\sigma}a^{}_{j\sigma}
  + U\sum_i n_{i\uparrow}n_{i\downarrow} ,
\label{generic}
\end{equation}
where $a^{\dagger}_{i\sigma}$ is a creation operator of an electron with spin 
$\sigma$ at site $i$, and $n_{i\sigma}=a^{\dagger}_{i\sigma}a^{}_{i\sigma}$. 
The hopping elements $t_{ij}=t$, $t^{\prime}$ and $t^{\prime\prime}$ stand
for the nearest neighbor, second-nearest neighbor, and third-neighbor hopping
on a 2D square lattice, and serve to model the electronic states of the
charge-transfer type in the cuprates. For convenience we choose $t$ as the 
energy unit.

It is interesting to note that hopping beyond nearest-neighbors contributes
to the energy and other properties not only in a 2D model, but also in the 
limit of $d\to\infty$. The energy contributions due to more distant neighbors
are finite due to the scaling of the hopping parameters on a hypercubic 
lattice. It is given by $t_{ij}\sim d^{-\|i-j\|/2}$ (see Refs.
\onlinecite{Met89,Vol93}), where $\|i-j\|$ is the distance between $i$ and 
$j$ defined by the 'bond metric', and gives the scaling factors 
$\sim 1/{\sqrt d}$ for first-, and $\sim 1/d$ for second- and third-nearest 
neighbor hopping, respectively, as the latter sites are two bonds apart. 

As mentioned above, we adopt the limit of infinite dimensions to determine
the spectral properties of the Hubbard Hamiltonian on a square lattice in 
the thermodynamic limit. In order to simplify the numerical evaluation of 
the self-energy we introduce an ansatz for the modified magnetic order in 
the doped systems, and assume incommensurate SS structures with a large but 
finite periodicity. This approach captures the essence of the competition 
between the weakened short-range AF order, and the kinetic energy induced by 
hole doping,\cite{Arr91} and allows to treat the systems in the thermodynamic 
limit at low temperature. The spiral states are characterized by the 
amplitude of the local magnetization, 
\begin{equation}
m_0=|\langle n_{i\uparrow}-n_{i\downarrow}\rangle| ,
\label{moment}
\end{equation}
which is independent of the site index $i$. The direction of the magnetic 
moment at each site $i$ is specified in the {\it global reference frame\/} 
by two spherical angles, $\Omega_i=(\phi_i,\theta_i)$, and therefore the 
original fermion operators, $\{a_{i\uparrow}^{\dagger},a_{i\downarrow}^
{\dagger}\}$, are transformed to the new fermions quantized with respect
to {\it the local quantization axis\/} at each site,\cite{Arr91}
\begin{equation}
c^{\dagger}_{i\sigma}=\sum_{\lambda}a^{\dagger}_{i\lambda}
   \left[{\cal R}(\Omega_i)\right]_{\lambda\sigma},
\label{rotation}
\end{equation}
where
${\cal R}(\Omega_i)=e^{-i(  \phi_i/2)\hat{\sigma}_z}
                    e^{-i(\theta_i/2)\hat{\sigma}_y}$ is the rotation matrix,
$\phi_i$ and $\theta_i$ are polar and azimuthal angle, respectively,
and $\hat{\sigma}_y$ and $\hat{\sigma}_z$ are Pauli spin matrices. This
transforms the Hubbard Hamiltonian (\ref{generic}) to the following form,
\begin{equation}
H\! =\! -\sum_{ij,\sigma\sigma^{\prime}}\!\!t_{ij}c^{\dagger}_{i\sigma}
\left[{\cal R}^{\dagger}(\Omega_i)
{\cal R}(\Omega_j)\right]_{\sigma\sigma^{\prime}}c^{}_{j\sigma^{\prime}}\! 
+ U\! \sum_i n_{i\uparrow}n_{i\downarrow}.
\label{hubbard}
\end{equation}
In the SS states we take the polar angle site-independent, $\theta_i=\theta$,
and the azimuthal angle is given by the wave-vector ${\bf Q}$ of the spiral, 
$\phi_i={\bf Q}\cdot{\bf R}_i$. Using the periodicity of the 
${\cal R}^{\dagger}{\cal R}$ matrix in Eq. (\ref{hubbard}), one finds after 
a Fourier transformation that the kinetic energy takes a simple $(2\times 2)$ 
matrix form,
\begin{eqnarray}
\hat{T}_{\bf Q}({\bf k})
&=&\case{1}{2}\;\varepsilon_{{\bf k}-\frac{\bf Q}{2}}
  (\hat{1}+\cos\theta\;\hat{\sigma}_z-\sin\theta\;\hat{\sigma}_x) \nonumber \\
&+&\case{1}{2}\;\varepsilon_{{\bf k}+\frac{\bf Q}{2}}
  (\hat{1}-\cos\theta\;\hat{\sigma}_z+\sin\theta\;\hat{\sigma}_x),
\label{ekin}
\end{eqnarray}
where $\varepsilon_{\bf k}=-2t(\cos k_x+\cos k_y)-4t'\cos k_x\cos k_y
-2t''(\cos 2k_x+\cos 2k_y)$ is the electron dispersion in a noninteracting
system in the global reference frame. Here we limit ourselves to plane 
spirals, and choose $\theta=\pi/2$. Therefore, the order parameter rotates 
in the $(a,b)$ plane, $\langle {\bf S}_i\rangle=
(m_0/2)[\cos({\bf Q}{\bf R}_i),\sin({\bf Q}{\bf R}_i),0]$. Double spirals
were shown to be unstable in the 2D $t$-$J$ model,\cite{Kan90} and we have no
reason to believe that they might be stabilized by further neighbor hopping.

In order to construct the leading local part of the self-energy, we use the 
DMFT and consider the impurity model coupled to the lattice by the effective
field (for more details see Ref. \onlinecite{Geo96}). The Anderson model of
a magnetic impurity coupled to a conduction band with SS order consists of
a ``non-degenerate impurity orbital'' at site $o$, with the fermion operators
$\{f_{o\sigma}^{\dagger},f_{o\sigma}^{}\}$, and the conduction electron bath
as an ``effective SS conduction band'' described by the operators
$\{c_{{\bf k}\sigma}^{\dagger},c_{{\bf k}\sigma^\prime}^{}\}$,
\begin{eqnarray}
&H_{imp}&=\varepsilon_{f}\sum_{\sigma} f_{o\sigma}^{\dagger} f_{o\sigma}^{} +
\sum_{{\bf k}\sigma \sigma^\prime}c_{{\bf k}\sigma}^{\dagger}
\left[ \tilde{T}_{\bf Q}({\bf k}) \right]_{\sigma \sigma^\prime}
c_{{\bf k}\sigma^\prime}^{}                                    \nonumber \\
&+& \sum_{{\bf k} \sigma \sigma^\prime} \left[ f_{o\sigma}^{\dagger}\,
[ V_{\bf Q}({\bf k})]_{\sigma\sigma^\prime}^{}c_{{\bf k}\sigma^\prime}^{}
+H.c. \right] + U n^f_{o\uparrow} n^f_{o\downarrow},
\label{anderson}
\end{eqnarray}
where $\varepsilon_{f}$ is an impurity energy level, 
and $\tilde{T}_{\bf Q}({\bf k})$ is an effective one-particle energy of the
same functional form as $\hat{T}_{\bf Q}({\bf k})$ (\ref{ekin}). The 
hybridization $(2\times 2)$ matrix in the local reference frame,
\begin{eqnarray}
\hat{V}_{\bf Q}({\bf k})
&=&\frac{1}{2}v_{{\bf k}-\frac{\bf Q}{2}}
  \left[\cos\frac{\theta}{2}(\hat{1}+\hat{\sigma}_z)-\sin\frac{\theta}{2}
  \hat{\sigma}_+\right] \nonumber \\
&+&\frac{1}{2}v_{{\bf k}+\frac{\bf Q}{2}}
  \left[ \cos\frac{\theta}{2} (\hat{1}-\hat{\sigma}_z) +
  \sin\frac{\theta}{2}\hat{\sigma}_-\right],
\label{hma}
\end{eqnarray}
where $\hat{\sigma}_{\pm}=\hat{\sigma}_x\pm i\hat{\sigma}_y$, is given by 
the individual hybridization elements in the global reference frame,
$v_{{\bf k}}=\sum_{i}e^{i{\bf k}\cdot {\bf R}_i}v_{oi}$. The Hamiltonian
(\ref{anderson}) is quadratic in $c_{{\bf k}\sigma}^{}$, and the bath of
conduction electrons can be integrated out giving raise to an effective
action of the impurity electrons which is of the usual form,\cite{Geo96}
\begin{eqnarray}
S_{eff}&=&-\sum_{\sigma \sigma'}\int_{0}^{\beta}d\tau d\tau'
\psi_{o\sigma}^{\star}(\tau)
{\cal G}^{0}_{{\bf Q}\sigma\sigma'}(\tau-\tau')^{-1}\psi_{o\sigma'}(\tau')
                                                    \nonumber \\
&+& U\int_{0}^{\beta}d\tau\; n^f_{o\uparrow}(\tau) n^f_{o\downarrow}(\tau),
\label{effaction}
\end{eqnarray}
where $\{\psi_{o\sigma}^{},\psi_{o\sigma^\prime}^{\star}\}$ are Grassmann
variables for the $f$-electrons. The {\em Weiss effective field\/}
${\cal G}^{0}_{{\bf Q}\sigma\sigma'}(\tau-\tau')$ is a $(2\times2)$
matrix in spin space,
\begin{eqnarray}
\hat{\cal G}^{0}_{\bf Q}(i\omega_{\nu})\,^{-1}&=&
i\omega_{\nu} - \varepsilon_f                        \nonumber \\
&-& \sum_{{\bf k}} {\hat V}_{\bf Q}^{}({\bf k})[i\omega_{\nu} 
- {\tilde T}_{\bf Q}({\bf k})]^{-1}{\hat V}_{\bf Q}^{\dagger}({\bf k}).
\label{impb}
\end{eqnarray}
For a plane spin spiral with $\theta=\pi/2$
the Weiss effective field becomes a diagonal matrix in spin space,
\begin{equation}
{\cal G}^{0}_{{\bf Q}\sigma \sigma'}(i\omega_n)
\sim \delta_{\sigma \sigma^\prime}.
\label{stabilityss}
\end{equation}
Note that this result only depends on the functional form of Eqs. 
(\ref{ekin})--(\ref{hma}), and not on the parameters, except that it holds
for a plane spiral. It implies that the local spin fluctuations are 
decoupled from the local charge fluctuations, and simplifies the present 
self-consistent calculation for SS states within the DMFT approach, as all 
local quantities including the self-energy $\Sigma$, are diagonal.

In the spirit of the DMFT approach, we approximate the Green function
using a {\it local self-energy},\cite{Met89,Geo96}
\begin{equation}
\hat{G}^{-1}_{\bf Q}({\bf k},i\omega_{\nu})=i\omega_{\nu}+\mu
- \hat{T}_{\bf Q}({\bf k}) - \hat{\Sigma}^{\rm HF}_{\bf Q}
- \hat{\Sigma}_{\bf Q}^{\rm SF}(i\omega_{\nu}),
\label{localg}
\end{equation}
where $\mu=-\varepsilon_f$ is the chemical potential.
The lattice (finite) dimensionality enters via the one-particle
energies $\hat{T}_{\bf Q}({\bf k})$ and gives rise to the
${\bf k}$-dependence of the spectral function. The lattice one-particle
Green function (\ref{localg}) is described by a $(2\times 2)$ matrix
$\hat{G}_{\bf Q}({\bf k},i\omega_{\nu})$ in spin space, where $\omega_{\nu}$
are fermionic Matsubara frequencies. The corresponding local lattice Green
function, $\hat{G}_{\bf Q}(i\omega_{\nu})=N^{-1}\sum_{\bf k}
\hat{G}_{\bf Q}({\bf k},i\omega_{\nu})\propto \delta_{\sigma\sigma'}$,
is diagonal in spin space due to the parity of the kinetic energy
$\hat{T}_{\bf Q}({\bf k})$. 

The self-energy consists of the HF part,
${\Sigma}^{\rm HF}_{{\bf Q}\sigma}=U\langle n_{0\bar{\sigma}}\rangle$
with $\bar{\sigma}=-\sigma$, and the spin-fluctuation (SF) part,
$\Sigma_{{\bf Q}\sigma}^{\rm SF}(i\omega_{\nu})$, which is determined by
the many-body effects. Using the {\it cavity method\/} \cite{Geo96} for a
hypercubic lattice at $d=\infty$, we verified that the dynamical Weiss field,
${\cal G}^{0}_{{\bf Q},\sigma}(i\omega_{\nu})$, can be computed from the
Dyson equation of the Anderson impurity model (\ref{anderson}) with 
broken spin-symmetry,
\begin{equation}
\hat{\cal G}^{0}_{\bf Q}(i\omega_{\nu})\,^{-1}=
\hat{G}_{\bf Q}(i\omega_{\nu})\,^{-1}
+\hat{\Sigma}_{\bf Q}^{\rm SF}(i\omega_{\nu})\;.
\label{cavity}
\end{equation}
Eqs. (\ref{localg}) and (\ref{cavity}) are fundamental in the
DMFT,\cite{Geo96} and can be solved self-consistently, provided an 
expression for the self-energy is known.

\subsection{Thermodynamic potential at finite temperature}
\label{sec:potential}

The calculations at finite temperature $T$ require the knowledge of the free
energy, $F(T,N_e)$, being a thermodynamic potential for a system of $N_e=Nn$
electrons.
It has to be minimized to find a stable SS state which determines the system
properties. The free energy may be found from the grand canonical potential,
$\Omega(T,\mu)$, using the standard approach for quantum many-body systems,
$F(T,N_e)=\Omega(T,\mu)+\mu N_e$.\cite{Bay62} For a translationally
invariant lattice model with {\em local self-energy} one finds the functional
form of the grand canonical potential,
\begin{eqnarray}
\Omega(T,\mu) &=&\Omega_{0}(T,\mu)+\Phi^{\rm SF}[{\hat G}_{\bf Q}]
                                                          \nonumber \\
&-& \beta^{-1} \sum_{{\bf k}\nu}
\ln\det \left[ 1 - \hat{G}^{0}_{\bf Q}({\bf k},i\omega_{\nu})
{\hat \Sigma}^{\rm SF}_{\bf Q}(i\omega_{\nu})\right]      \nonumber \\
&-& \beta^{-1}N \sum_{\nu}Tr\left[{\hat\Sigma}^{\rm SF}_{\bf Q}(i\omega_{\nu})
{\hat G}_{\bf Q}(i\omega_{\nu})\right],
\label{kbpo}
\end{eqnarray}
with $\beta=1/k_BT$, and ${\hat\Sigma}^{\rm SF}_{\bf Q}(i\omega_{\nu})$ is
the self-energy discussed below. The functional (\ref{kbpo}) is stationary,
i.e., $\delta\Omega=0$ ensures that the minimum of the grand canonical
potential has been found, and determines the self-energy from the
Luttinger-Ward functional,
\begin{equation}
\Sigma^{\rm SF}_{\sigma\sigma}[{\hat G}_{\bf Q}]=\frac{1}{N}\;
\frac{\delta \Phi^{\rm SF}[{\hat G}_{\bf Q}]}
{\delta {\hat G}_{{\bf Q},\sigma\sigma}}.
\label{luwa}
\end{equation}
Our perturbative expansion is constructed around the symmetry-broken HF state,
hence the grand canonical potential of the ``non-interacting'' reference
system includes a correction term to avoid double-counting and reads,
\begin{eqnarray}
\Omega_{0}(T,\mu) &=& \beta^{-1} \sum_{{\bf k}\nu}
\ln \det \left[{\hat G}^{0}_{\bf Q}({\bf k},i\omega_{\nu})\right] \nonumber \\
&-& UN \langle n_{o \uparrow} \rangle \langle n_{o \downarrow} \rangle.
\label{kbpoHF}
\end{eqnarray}
The spectrum which defines $\Omega_{0}(T,\mu)$ is given by the Green's
function in the HF approximation,
\begin{equation}
\hat{G}^{0}_{\bf Q}({\bf k},i\omega_{\nu})^{-1}=i\omega_{\nu}+\mu
- \hat{T}_{\bf Q}({\bf k}) - \hat{\Sigma}^{\rm HF}_{\bf Q}.
\label{greenHF}
\end{equation}

The Luttinger-Ward functional $\Phi^{\rm SF}[G_{\bf Q}]$ in Eq. (\ref{kbpo})
is defined via the diagrammatic expansion of $\Sigma^{\rm SF}_{\bf Q}$ in
terms of the full Green's function $G_{\bf Q}$. The self-energy of the
infinite-dimensional Hubbard model is a local dynamical quantity and involves
only the local component of the Green's function (\ref{localg}). This implies
that $\Phi^{\rm SF}[G]=N\Phi^{\rm SF}_{\rm imp}[G]$,\cite{Jan92} meaning that
the functional $\Phi^{\rm SF}[G]$ can be approximated by some infinite subset
of the one-particle irreducible closed Feynman diagrams of the Anderson
impurity model (\ref{anderson}). We take for $\Phi^{\rm SF}[G]$ the sum of 
all particle-hole diagrams,\cite{Bic89} and the effective particle-hole 
interaction $\bar U$,\cite{Che91,Fle97}
\begin{eqnarray}
\Phi^{\rm SF} &=& N(\phi_{2} + \phi_{\perp} + \phi_{\|}),     \\
\phi_{2} & = & -\frac{1}{2} \beta^{-1}\bar{U}^{2}
\sum_{\mu} \chi^{(0)}_{\uparrow\uparrow}(i\omega_{\mu})
\chi^{(0)}_{\downarrow\downarrow}(i\omega_{\mu}), \\
\phi_{\perp} & = & \beta^{-1}\sum_{\mu} \ln
\left[1-\bar{U} \chi^{(0)}_{\uparrow\downarrow}(i\omega_{\mu}) \right] +
\beta^{-1}\bar{U} \sum_{\mu} \chi^{(0)}_{\uparrow\downarrow}(i\omega_{\mu})
\nonumber \\&+&
\frac{1}{2}\beta^{-1} \bar{U}^{2} \sum_{\mu}
\chi^{(0)}_{\uparrow\downarrow}(i\omega_{\mu})
\chi^{(0)}_{\downarrow\uparrow}(i\omega_{\mu}), \\
\phi_{\|} & = & \frac{1}{2}\beta^{-1} \sum_{\mu} \ln
\left[1-\bar{U}^{2}
\chi^{(0)}_{\uparrow\uparrow}(i\omega_{\mu})
\chi^{(0)}_{\downarrow\downarrow}(i\omega_{\mu}) \right] \nonumber\\
&+& \frac{1}{2}\beta^{-1} \bar{U}^{2} \sum_{\mu}
\chi^{(0)}_{\uparrow\uparrow}(i\omega_{\mu})
\chi^{(0)}_{\downarrow\downarrow}(i\omega_{\mu}),
\label{kbdia}
\end{eqnarray}
where
\begin{equation}
\chi_{\sigma\sigma^{\prime}}^{(0)}(i\omega_{\mu})=-\beta^{-1}
\sum_{\nu}{\cal G}^{0}_{{\bf Q},\sigma}(i\omega_{\nu})\,
          {\cal G}^{0}_{{\bf Q},\sigma^{\prime}}(i\omega_{\nu}+i\omega_{\mu}),
\label{nsus}
\end{equation}
is the noninteracting particle-hole susceptibility. Self-consistency would
require that $\Phi^{\rm SF}=\Phi^{\rm SF}[G_{\bf Q}]$; here instead we apply
the non-self-consistent procedure introduced by Bulut, Scalapino, and
White,\cite{Bul93} and approximate
$\Phi^{\rm SF}[G_{\bf Q}]\to\Phi^{\rm SF}[{\cal G}^{0}_{\bf Q}]$. It has been
shown that this procedure may be regarded to be a reasonable approximation
as the thermodynamic potential (\ref{kbpo}) is stationary and one expects not
to move too far away from its minimum.

\subsection{Self-energy with local spin fluctuation}
\label{sec:sigma}

It is known to be notoriously difficult to obtain an analytic expression 
for the self-energy, and so far mostly an ansatz within the iterative 
perturbation scheme (IPS) based on SOPT has been used.\cite{Kaj96} The ansatz 
introduces an approximate form of self-energy which starts from the SOPT and
allows to reproduce the exact results in certain limits. Although this 
approach reproduces the correct large $U$ limit,\cite{Kaj96} it overestimates 
the correlation effects in the nonmagnetic states, and thus becomes 
uncontrollable in the intermediate $U$ regime.
Therefore, it cannot be applied to investigate the phase stability and
dynamics in the magnetic states of the Hubbard model. We have verified that
the AF LRO disappears in the 2D Hubbard model ($t'=t''=0$) at half-filling
for $U\simeq 5t$ for $t'=t''=0$, if the formula introduced by Kajueter and
Kotliar\cite{Kaj96} is used (see Sec. \ref{sec:mott}).

The SF part of the self-energy,
$\Sigma_{{\bf Q}\sigma}^{\rm SF}(i\omega_{\nu})$ follows from the
Kadanoff-Baym potential (\ref{kbpo}) containing a class of diagrams up to
infinite order,
\begin{eqnarray}
\Sigma_{{\bf Q}\sigma}^{\rm SF}(i\omega_{\nu})
&=&\frac{{\bar U}^{2}}{\beta} \sum_{\mu}
\chi_{\bar{\sigma}\sigma{\bf Q}}(i\omega_{\mu})\,
{\cal G}^{0}_{{\bf Q}\bar{\sigma}}(i\omega_{\nu}-i\omega_{\mu}) \nonumber \\
 & + & \frac{{\bar U}^{2}}{\beta} \sum_{\mu}
\chi_{\bar{\sigma}\bar{\sigma},{\bf Q}}(i\omega_{\mu})\,
{\cal G}^{0}_{{\bf Q}\sigma}(i\omega_{\nu}-i\omega_{\mu}).
\label{sigma}
\end{eqnarray}
Here we approximated $\Sigma[G]$ by $\Sigma[{\cal G}^0]$ and avoid
self-consistency. The transverse part in Eq. (\ref{sigma}) resembles the
self-energy derived by the coupling of the moving hole to transverse
spin-fluctuations, as derived using the spin-wave theory.\cite{Alt95}
However, the longitudinal part is not included in the latter approach,
and we find that it cannot be neglected in the relevant regime of
parameters for high temperature superconductors.

The self-energy in a magnetic system is calculated using the Weiss effective
field (\ref{cavity}) in the symmetry-broken magnetic state. The transverse,
\begin{equation}
\chi_{\bar{\sigma}\sigma}(i\omega_{\mu})=
{\bar U}\frac{[\chi^{(0)}_{\bar{\sigma}\sigma}(i\omega_{\mu})]^2}
{1-{\bar U}\chi^{(0)}_{\bar{\sigma}\sigma}(i\omega_{\mu})},
\label{trans}
\end{equation} 
and longitudinal,
\begin{equation}
\chi_{\sigma\sigma}(i\omega_{\mu})=
\frac{\chi^{(0)}_{\sigma\sigma}(i\omega_{\mu})}{1-{\bar U}^2
 \chi^{(0)}_{\uparrow\uparrow}(i\omega_{\mu})
 \chi^{(0)}_{\downarrow\downarrow}(i\omega_{\mu})},
\label{longi}
\end{equation}
susceptibility in Eq. (\ref{sigma}) are found in random phase approximation
(RPA) with renormalized interaction $\bar{U}$. Here the non-interacting
susceptibilities, $\chi^{(0)}_{\sigma\sigma}(i\omega_{\mu})$, are
calculated from the dynamical Weiss field Green function (\ref{cavity}).

We would like to emphasize that the renormalized interaction $\bar U$ is not
a fitting parameter,\cite{Bul93} but results from the static screening by
particle-particle diagrams which leads to\cite{Che91,Fle97}
\begin{equation}
\label{ubar}
\bar{U}=U/[1+U\chi^{pp}(0)],
\end{equation}
where the particle-particle vertex is again determined by the Weiss field,
\begin{equation}
\chi^{pp}(0)=\beta^{-1}
\sum_{\mu}{\cal G}^{0}_{{\bf Q}  \uparrow}( i\omega_{\mu})\,
          {\cal G}^{0}_{{\bf Q}\downarrow}(-i\omega_{\mu}) .
\label{kern}
\end{equation}
This screening effect gives the magnetic structure factor \cite{Che91} and
the self-energy \cite{Bul93} calculated from Eq. (\ref{sigma}) in good
agreement with the QMC results, and depends on the underlying magnetic order.
It is largest in the paramagnetic states and vanishes in the N\'eel state at
$n=1$ for $U\to\infty$, and is thus very important when the magnetic phase
diagrams are considered.\cite{Fle97} The proposed self-energy (\ref{sigma})
expresses the spin-fluctuation exchange interaction \cite{Ber66} with an
effective potential due to particle-particle scattering.\cite{Che91}

Eqs. (\ref{localg}), (\ref{cavity}) and (\ref{sigma}) represent a solution
for the one-particle Green function within the DMFT. They have been solved 
self-consistently and the energetically stable spiral configuration was found. 
This procedure is further justified by the sum rule,\cite{Vil97}
\begin{equation}
\label{sumrule}
\frac{1}{2\beta}\sum_{\nu\sigma}
     \Sigma_{\sigma}(i\omega_{\nu})G_{\sigma}(i\omega_{\nu})
     e^{i\omega_{\nu}0^+}
     =U\langle n_{0\uparrow}n_{0\downarrow}\rangle,
\end{equation}
which is well fulfilled in the present approach with
$U\langle n_{0\uparrow}n_{0\downarrow}\rangle\simeq \bar{U}
  \langle n_{0\uparrow}\rangle\langle n_{0\downarrow}\rangle$.\cite{Fle97}
We also show below (Sec. \ref{sec:mott}) that the local self-energy
(\ref{sigma}) leads to an overall satisfactory agreement with the QMC and
ED data.

\section{Excitation spectra}
\label{sec:exci}

\subsection{Photoemission at finite temperature}
\label{sec:pes}

A complete theory of photoemission (PES) would require an analysis not only
of the one-particle Green's function but also of the three-particle Green's
functions. We would like to point out that quantitative calculations of the
three-particle Green's function for strongly correlated systems has not yet
proven feasible. However, some aspects of the problem can be discussed in
terms of the one-electron spectrum, provided that ``final-state'' or
``particle-hole'' interactions can be neglected. Under this assumption the
problem simplifies and the PES spectrum may be determined using the
one-particle Green function alone. Such an approach which neglects
final-state and particle-hole interactions has been applied with success to
interpret \cite{Bal95} the dispersion found in the ARPES data of the 
copper oxides.\cite{Mar96,Kim98}

Here we shall derive the relation of the PES spectra to the
one-electron spectral function within the ``sudden'' approximation, where
final state interactions are neglected.\cite{Hed69}. To be specific, let us
consider a transition from a state $|n\rangle$ with energy $E_n$ into a state
of the form $A_{{\bf k}\nu}^{\dagger}|m\rangle$, in which we treat the
photoelectron in state $|{\bf k}\nu\rangle$ as dynamically decoupled, but
retain the full many-body interactions of the electrons in the bulk described
by the Hubbard model Hamiltonian. The PES intensity for the
magnetic system with an incommensurate magnetic order is nontrivial within
the DMFT approach, as one cannot use a Bogoliubov transformation to establish
the relation between the measured electrons and their local states in the SS
state. The outgoing photoelectron is observed in the {\it global reference
system\/}, whereas the quantum states of the bulk $|n\rangle$ have to be
considered within the {\it local reference system\/} for the spin degrees of
freedom. For clarity we write in the following the operators for the
scattered states in capital letters and the operators for the electronic
states of the solid described by the present model Hamiltonian in lower-case
letters.

At finite temperature $T$ we consider the probability density of the
absorption of a photon with frequency $\omega$ in a grand canonical ensemble
and obtain,
\begin{equation}
{\mathcal W}({\bf k},\omega) = Z^{-1} \sum_{mn} e^{-\beta K_n}
\left| I_{m n}^{\nu} \right|^2 \delta(\varepsilon_{{\bf k}}+K_m-K_n-\omega),
\label{prophoto}
\end{equation}
with $K_n \equiv E_n - \mu N_n$ and the partition function
$Z=\sum_{n} e^{-\beta K_n}$. The amplitude of the transition,
\begin{equation}
I_{m n}^{\nu} =
\langle m | A_{{\bf k}\nu}
\sum_{{\bf p} {\bf p}^\prime \atop \sigma\sigma^\prime}
A_{{\bf p}\sigma}^{\dagger}
[\Delta_{{\bf Q}}({\bf p},{\bf p}^\prime)]_{\sigma\sigma^\prime}
c_{{\bf p}^\prime-\frac{\bf Q}{2} \sigma^\prime}^{} | n \rangle
\label{matphoto}
\end{equation}
is determined by the optical matrix element,
\begin{eqnarray}
\hat{\Delta}_{\bf Q}({\bf p},{\bf p}^\prime)
&=&\frac{1}{2}\Delta_{{\bf p},{\bf p}^\prime}
  [\cos\frac{\theta}{2}(\hat{1}-\hat{\sigma}_z)
  +\sin\frac{\theta}{2} \hat{\sigma}_-]                  \nonumber \\
&+&\frac{1}{2}\Delta_{{\bf p},{\bf p}^\prime-{\bf Q}}
  [\cos\frac{\theta}{2}(\hat{1}+\hat{\sigma}_z)
  -\sin\frac{\theta}{2} \hat{\sigma}_+] ,
\label{oma}
\end{eqnarray}
and can be calculated using the Bloch wave-functions in the
{\it global reference system\/},
\begin{equation}
\Delta_{{\bf p},{\bf p}^\prime}=\frac{1}{2}\sum_{\sigma}
\langle \Psi_{{\bf p}\sigma}| \mbox{\boldmath$\epsilon$} \cdot {\bf k}
|\psi_{{\bf p}^{\prime}\sigma}\rangle,
\label{omal}
\end{equation}
where \mbox{\boldmath$\epsilon$} is the polarization vector. The operator
${\bf k}=-i\mbox{\boldmath$\nabla$}$ conserves total momentum in the 
scattering plane, so that $\Delta_{{\bf p},{\bf p}^\prime}\propto
\delta_{{\bf p}_{\parallel},{\bf p}^\prime+{\bf K}}$, where 
${\bf K}$ is a 2D lattice vector, and ${\bf p}_{\parallel}$ is the 
photoelectron momentum component in the 2D plane.

For solids the outgoing wave solution is the ``time-inverted low-energy
electron diffraction (LEED) state''.\cite{Bar85} The LEED state consists of
an incoming plane wave, reflected plane waves, and a combination of Bloch
waves inside the solid which fulfill the matching boundary conditions. In
lowest order we have one (damped) Bloch wave travelling away from the
surface. In the time inverted (complex conjugated) state the Bloch wave
travels towards the surface, and goes over in a plane wave outside. The LEED
scattered waves become incoming waves on time inversion, and give no
contribution to the photocurrent. The photoelectron is usually detected at
energies which are much higher than the typical energy regime described by
the Hubbard model, and therefore the Bloch waves
occupy high-energy quantum states which are initially unoccupied,
\begin{equation}
A_{{\bf k}\sigma}^{}\, |n\rangle \simeq 0 .
\label{sudden}
\end{equation}
Hence, we obtain for the plane SS state ($\theta=\pi/2$),
\begin{eqnarray}
{\mathcal W}({\bf k},\omega) &=& - \frac{1}{\pi}
\sum_{\sigma\sigma^\prime} |\Delta_{{\bf k},{\bf k}}|^{2}
n_{\rm F}(\epsilon_{{\bf k}} - \omega)                      \nonumber \\
& & \hskip 1cm \times {\rm Im} G_{\sigma \sigma^\prime}
({\bf k}-{\bf Q}/2,\varepsilon_{{\bf k}}-\omega) ,
\label{prophotof}
\end{eqnarray}
where $n_{\rm F}(\omega)$ is the Fermi function, and the following identity,
valid only for plane spirals ($\theta=\pi/2$),
has been used ($\lambda_{\sigma}=1,-1$ for $\sigma=\uparrow,\downarrow$),
\begin{equation}
\sum_{\sigma\sigma^\prime}
G_{\sigma \sigma^\prime}({\bf k}-{\bf Q}/2,\omega) =
\sum_{\sigma \sigma^\prime} \lambda_{\sigma} \lambda_{\sigma^{\prime}}
G_{\sigma \sigma^\prime}({\bf k}+{\bf Q}/2,\omega) .
\label{idgreen}
\end{equation}

Within the ``sudden'' approximation the measured PES spectra near the Fermi 
energy can therefore be related to the one-electron spectral function Eq. 
(\ref{prophotof}) of the system with local spin-quantization axes, defined by
\begin{eqnarray}
{\rm Im} G_{\sigma \sigma^\prime}({\bf k},\omega)&=&
-\frac{\pi}{Zn_{\rm F}(\omega)}\sum_{mn} 
\langle n|c_{{\bf k}\sigma^\prime}^{\dagger}|m\rangle 
\langle m|c_{{\bf k}\sigma}^{}|n\rangle                    \nonumber \\
& & \hskip 1.0cm \times
e^{-\beta K_n}\delta(\omega-K_n+K_m).
\label{spectraT}
\end{eqnarray}
Therefore, the total
one-particle excitation spectra is described by the spectral function,
\begin{equation}
A({\bf k},\omega)=-\frac{1}{\pi}\sum_{\sigma\sigma'}{\rm Im}\;
G_{{\bf Q},\sigma\sigma'}\left({\bf k}-\frac{\bf Q}{2},\omega+i\epsilon\right),
\label{apes}
\end{equation}
where $G_{{\bf Q},\sigma\sigma'}({\bf k}-{\bf Q}/2,\omega+i\epsilon)$ is 
given by Eq. (\ref{localg}), and a numerical broadening $\epsilon>0$.
The electron occupation number $\langle n_{\bf k}\rangle$ normalized per one
spin, equal to the one-electron removal sum, can be obtained without analytic
continuation of the Matsubara Green's function (\ref{localg}) by performing
a direct summation over the Matsubara frequencies,
\begin{equation}
\langle n_{\bf k} \rangle = \frac{1}{2\beta}\sum_{\nu,\sigma\sigma'}
e^{i \omega_{\nu}0^+}
G_{{\bf Q},\sigma\sigma'}\left({\bf k}-\frac{\bf Q}{2}, i\omega_{\nu}\right).
\label{momdis}
\end{equation}
Finally, we calculate also the total densities of states in the AF and SS
states using the derived spectral functions (\ref{apes}),
\begin{equation}
N(\omega)=\frac{1}{N}\sum_{\bf k}A({\bf k},\omega).
\label{dos}
\end{equation}

\subsection{Optical conductivity }
\label{sec:transport}

We derived the complex optical conductivity $\sigma_{xx}(\omega)$ for the
spiral magnetic order following the formalism introduced by Shastry and
Sutherland,\cite{Sha90} and by Scalapino, White and Zhang.\cite{Sca93}
Their derivation has to be generalized to the case of extended hopping.
Moreover, as the symmetry is locally broken in a magnetic system with local
quantization axes, the calculation of the optical conductivity is not
straightforward. The Hubbard Hamiltonian (\ref{hubbard}) within the local
reference system for the spin quantization axis and first-, second-, and
third-nearest neighbor hopping elements, $t_{il}=t,t^\prime,t^{\prime\prime}$,
respectively, has an electron kinetic energy,
\begin{eqnarray}
K=-\left.\sum_{il,\sigma\sigma^{\prime}}\right.^{\prime} &t_{il}&
\left[ c^{\dagger}_{i\sigma}\left[{\cal R}^{\dagger}(\Omega_i)
{\cal R}(\Omega_l)\right]_{\sigma\sigma^{\prime}}
c^{}_{l\sigma^{\prime}}\right.                       \nonumber \\
&+& \left. c^{\dagger}_{l\sigma^{\prime}}\left[{\cal R}^{\dagger}(\Omega_l)
{\cal R}(\Omega_i)\right]_{\sigma^{\prime}\sigma}c^{}_{i\sigma}\right],
\label{kinpart}
\end{eqnarray}
where $\left.\sum_{il}\right.^{\prime}$ indicates a restricted sum,
with ${\bf R}_{l}={\bf R}_{i}+\| i-l\|_x{\bf x}+\| i-l\|_y{\bf y}$ around
a given lattice site $i$, and ${\bf x}=(1,0)$, ${\bf y}=(0,1)$ are unit 
lattice vectors. We introduce a directed ``bond metric'', 
$\| i-l\|_{x(y)}$, which is a distance between two sites, $i$ and $l$, {\it 
on the lattice\/}, and counts the number of $x(y)-$oriented bonds that 
connect site $i$ with site $l$, respectively, e.g., $\| i-l\|_x=2$ and 
$\| i-l\|_y=0$ if the electron hops to a third-nearest neighbor with 
amplitude $t^{\prime\prime}$ along an $x-$oriented link. Here 
${\cal R}(\Omega_i)$ is the unitary matrix which transforms the original 
fermions, $\{a_{i\uparrow}^{\dagger},a_{i\downarrow}^{\dagger}\}$, into the 
fermions quantized with respect to local quantization axes at each site,
$\{c_{i\uparrow}^{\dagger},c_{i\downarrow}^{\dagger}\}$, introduced in Eq.
(\ref{rotation}). In what
follows we are interested in the current response to a vector potential along
the $x$-direction of the 2D square lattice $A_x(l,t)$. In the presence of a
vector potential, the hopping term is modified by the Peierls phase
factor,\cite{Sca93} either
$\exp\{+ieA_x(l,t)\| i-l\|_x\}$ or
$\exp\{-ieA_x(l,t)\| i-l\|_x\}$, for $t_{il}$ or $t_{li}$, respectively.
Expanding these phase factors in the usual manner up to second order
$\sim A^2$ one finds,
\begin{equation}
K_{A}=K-\left.\sum_{il}\right.^\prime \left[
e j_{x}^{P}(i-l) A_{x}(l) + \frac{e^{2}}{2} k_{x}(i-l)A_{x}(l)^2 \right].
\label{kinpotpart}
\end{equation}
Here $j_x^P(i-l)$ is the $x$-component of the paramagnetic current density,
\begin{eqnarray}
j_{x}^{P}(i-l)=i \sum_{\sigma\sigma^{\prime}} \| i&-&l\|_x t_{il}
\left[ c^{\dagger}_{i\sigma}\left[{\cal R}^{\dagger}(\Omega_i)
{\cal R}(\Omega_l)\right]_{\sigma\sigma^{\prime}}
c^{}_{l\sigma^{\prime}}\right.                       \nonumber \\
&-& \left. c^{\dagger}_{l\sigma^{\prime}}\left[{\cal R}^{\dagger}(\Omega_l)
{\cal R}(\Omega_i)\right]_{\sigma^{\prime}\sigma}c^{}_{i\sigma}\right],
\label{parcurdens}
\end{eqnarray}
and $k_{x}(i-l)$ is the kinetic-energy contribution due to the $x-$oriented
links, weighted by the metric factor connecting site $i$ with site $l$,
\begin{eqnarray}
k_x(i-l) = - \sum_{\sigma\sigma^{\prime}} \| i&-&l\|_x^2 t_{il}
\left[ c^{\dagger}_{i\sigma}
\left[{\cal R}^*(\Omega_i){\cal R}(\Omega_l)\right]_{\sigma\sigma^{\prime}}
c^{}_{l\sigma^{\prime}} \right.                      \nonumber \\
&+& \left. c^{\dagger}_{l\sigma^{\prime}}
\left[{\cal R}^*(\Omega_l){\cal R}(\Omega_i)\right]_{\sigma^{\prime}\sigma}
c^{}_{i\sigma}\right],
\label{kinx}
\end{eqnarray}

After performing a Fourier transformation one finds the average contribution
of kinetic energy (\ref{kinx}) per one site,
\begin{equation}
\langle k_x\rangle=
\frac{1}{N}\sum_{{\bf k},\sigma\sigma'}\langle c^{\dagger}_{{\bf k}\sigma}
[\hat{t}_{x,{\bf Q}}({\bf k})]^{}_{\sigma\sigma'}c^{}_{{\bf k}\sigma'}\rangle,
\label{xkin}
\end{equation}
with the coupling between the transformed elements at momenta
${\bf k}-{\bf Q}/2$ and ${\bf k}+{\bf Q}/2$ due to the magnetic order,
\begin{eqnarray}
\hat{t}_{x,{\bf Q}}({\bf k})
&=&\frac{1}{2}\varepsilon_x\!\left({\bf k}-\frac{\bf Q}{2}\right)\!
  \left(\hat{\openone}+\hat{\sigma}_z\cos\theta-\hat{\sigma}_x\sin\theta\right)
                                                       \nonumber \\
&+&\frac{1}{2}\varepsilon_x\!\left({\bf k}+\frac{\bf Q}{2}\right)\!
  \left(\hat{\openone}-\hat{\sigma}_z\cos\theta+\hat{\sigma}_x\sin\theta\right),
\label{ekinx}
\end{eqnarray}
and $\varepsilon_x({\bf k})=-2t\cos k_x-4t'\cos k_x\cos k_y-8t''\cos 2k_x$.

As usually, the optical conductivity in long-wavelength limit ${\bf q}\to 0$,
$\sigma_{xx}(\omega)=\sigma_{xx}'(\omega)+i\sigma_{xx}''(\omega)$,
is determined by the current response to a vector potential along the
$x$-direction,\cite{Sca93} and one finds using the Kubo linear response 
theory,
\begin{equation}
\sigma_{xx}(\omega) = - e^2 \frac{\langle -k_{x} \rangle -
\Lambda_{xx}({\bf q}={\bf 0},\omega+i0^+)}{i(\omega+i0^+)},
\label{optcon}
\end{equation}
where $\Lambda_{xx}({\bf q},i\omega_{\mu})$ is the current-current
correlation function,
\begin{equation}
\Lambda_{xx}({\bf q},i\omega_{\mu})=\frac{1}{N}\int_{0}^{\beta}d\tau
e^{i\omega_{\mu}\tau}\langle j_x({\bf q},\tau) j_x(-{\bf q},0)\rangle.
\label{currcorr}
\end{equation}
The latter correlation function is given exactly by the particle-hole bubble
diagram,\cite{Khu90,Pru95} where for ${\bf q}\to 0$,
\begin{equation}
j_{x}=\sum_{{\bf k},\sigma\sigma'}c^{\dagger}_{{\bf k}\sigma}
[\hat{j}_{x,{\bf Q}}({\bf k})]_{\sigma \sigma'} c^{}_{{\bf k}\sigma'},
\label{curr}
\end{equation}
and for the present SS state,
\begin{eqnarray}
\hat{j}_{x,{\bf Q}}({\bf k})
&=&\frac{1}{2}\;j_x\!\left({\bf k}-\frac{\bf Q}{2}\right)
  (\hat{1}+\hat{\sigma}_z\cos\theta-\hat{\sigma}_x\sin\theta) \nonumber \\
&+&\frac{1}{2}\;j_x\!\left({\bf k}+\frac{\bf Q}{2}\right)
  (\hat{1}-\hat{\sigma}_z\cos\theta+\hat{\sigma}_x\sin\theta),
\label{jx}
\end{eqnarray}
with $j_x({\bf k})=2t\sin k_x+4t'\sin k_x\cos k_y+4t''\sin 2k_x$. The
advantage of using the DMFT with the local self-energy is that the vertex
corrections to the current-current correlation function (\ref{currcorr})
disappear, and the optical conductivity can be calculated without further
approximations.\cite{Khu90}

We have verified for large variety of doping levels and temperatures that 
the following optical sum rule,
\begin{equation}
2\int_{0}^{\infty}d\omega\, \sigma_{xx}'(\omega)=e^2\pi\;\langle -k_x\rangle ,
\label{sumo}
\end{equation}
is always fulfilled within the numerical accuracy, in contrast to the 
approaches which cannot be derived in a diagrammatic way.
Eq. (\ref{sumo}) is also used to define the plasma-frequency $\omega_p$,
\begin{equation}
\omega_p^2=8\int_{0}^{\infty} d\omega \sigma_{xx}'(\omega).
\label{omegap}
\end{equation}
For the discussion of the complex conductivity function, it is convenient to
introduce the following parametrization by the scattering rate,
$\tau(\omega)^{-1}$, and the effective mass $m^*(\omega)/m_e$ ($m_e$ is the
electron mass),\cite{Uch91}
\begin{equation}
\sigma_{xx}(\omega)=\frac{\omega_p^2}{4\pi}\;
\frac{1}{\tau^{-1}(\omega) - i\omega \frac{m^*(\omega)}{m_e}}.
\label{param}
\end{equation}

>From the real part of the optical conductivity (\ref{optcon}) we find in
the limit $\omega\to 0$ the static conductivity,
\begin{equation}
\sigma_{xx}'(\omega=0)=e^2\pi D+e^2 \,\lim_{\omega\to 0}\frac{1}{\omega}
{\rm Im}\,\Lambda_{xx}({\bf q}=0,\omega),
\label{dcc}
\end{equation}
with the Drude weight $D$ which may be obtained from the zero-temperature
extrapolation of the current-current correlation function in the upper
complex plane,\cite{Sca93}
\begin{equation}
D = \lim_{T\to 0}\left[\langle -k_x\rangle -
{\rm Re}\; \Lambda_{xx}({\bf q}=0,2\pi iT)\right] .
\label{Druden}
\end{equation}
The optical conductivity allows to determine the in-plane static resistivity,
\begin{equation}
\rho_{xx}(T) = \sigma_{xx}'(\omega=0,T)^{-1},
\label{dcr}
\end{equation}
where the static conductivity $\sigma_{xx}'(\omega=0,T)^{-1}$ is obtained as
in Eq. (\ref{dcc}). We present the results obtained for the optical
conductivity and static resistivity in Sec. \ref{sec:optic}, and show that
the magnetic order in the doped compounds has directly measurable 
consequences for these quantities.

\section{ One-particle spectra }
\label{sec:results}

\subsection{Quasiparticles at half filling}
\label{sec:mott}

The ground state of the Hubbard model with nearest neighbor hopping
($t'=t''=0$) on a square lattice is an AF insulator. The insulating behavior
and the gap develop gradually at half-filling with increasing $U$ starting
from $U=0$ due to the perfect nesting instability, leading to a Slater gap.
This gap changes into a Mott-Hubbard gap under increasing $U$, and the system
approaches the limit of a Heisenberg antiferromagnet.\cite{Cha77} This regime
of large $U$ was found to be difficult for a quantitative description within
the DMFT approches,\cite{Geo96} as an accurate determination of the energy
gains due to AF long-range order is there of crucial importance. Therefore,
the attempts to describe the AF order based on the SOPT within the IPS
failed and the magnetic order disappeared at larger $U$.\cite{Kaj96}
In contrast, the QMC calculation in the $d\to\infty$ limit gave a stable
AF state for large $U>4t$.\cite{Jar92,Ulm95}

\begin{figure}
\centerline{\psfig{figure=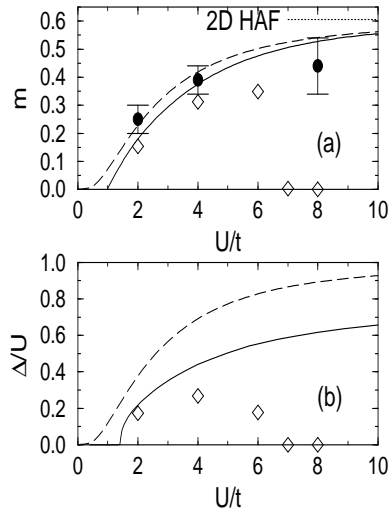,height=2.8in,width=2.0in}}
\narrowtext
\smallskip
\caption
{Antiferromagnetic state for the Hubbard model at $n=1$:
 (a) magnetization $m$ (\protect\ref{magnet}), and
 (b) the AF gap $\Delta/U$ in the 2D Hubbard model,
 as obtained using the HF approximation at $T=0$ (dashed lines) and
                   DMFT approach at $T=0.05t$ (full lines).
 The data points in (a) are QMC results reproduced from Ref.
 \protect\onlinecite{Hir89}. The diamonds 
 show the results of the IPS with the self-energy calculated in SOPT. }
\label{fig:magnet}
\end{figure}

We treat here the range of large $U\simeq W$ ($W=8t$) as a test case for our
analytic method. The calculations were performed at a low temperature
$T=0.05t$ which allows to describe the magnetic excitations ($T\ll J=4t^2/U$).
They gave an AF ground state at $n=1$ which reproduces correctly the
localization of electrons in the limit of $U\to\infty$. The magnetization
$m_0$ (\ref{moment}) is only slightly reduced by the dynamical effects with
respect to its HF value, and approaches the HF limit at $U\to\infty$. The
ground state is the N\'eel AF state, as found in the $d\to\infty$ 
limit.\cite{Ken88} Thus, we reduced the self-consistently obtained values of 
the mean-field magnetization $m_0$ (\ref{moment}) by a factor 0.606 in order 
to simulate the known reduction of $m_0$ by intersite quantum fluctuations 
in a 2D lattice,\cite{Zla90}
\begin{equation}
m = 0.606m_0 .
\label{magnet}
\end{equation}
After this reduction the calculated values of $m$ approach the value of 
0.606 in the limit $U\to\infty$ [Fig. \ref{fig:magnet}(a)]. One finds also 
a very good agreement with the QMC data \cite{Hir89} at $U/t=2$ and 4, and 
a reasonable agreement at $U/t=8$.

In contrast, the AF gap $\Delta$ is significantly reduced from its HF value
[Fig. \ref{fig:magnet}(b)]. This reduction follows from a drastic change of
the one-particle spectra by dynamical effects which lead to QP states at the 
edge of the Mott-Hubbard gap which are accompanied by a large incoherent part 
at higher energies. Also the reduction of the gap found in ED,\cite{Dag92} 
comes out correctly as shown in Ref. \onlinecite{Fle98}. For example, we 
have found a gap of $4.93t$ at $U/t=8$, while the corresponding gap in the 
HF calculation is $7.14t$. This gap reduction can also be captured by the
leading dynamical correlations described within the SOPT, but only in the
regime of $U<2.5t$. The discrepancy between the SOPT and the DMFT results
increases with increasing $U$, with the gap and the magnetization $m$ being
too small, and finally the AF order disappears and the gap closes at
$U\simeq 7t$. This shows a very limited applicability of the approaches 
using the self-energy based on the SOPT,\cite{Kaj96} which are known to 
underestimate the region of stability of magnetic states and fail at large 
$U$ due to the uncontrolled increase of the correlation energy in 
nonmagnetic states.

\begin{figure}
\centerline{\psfig{figure=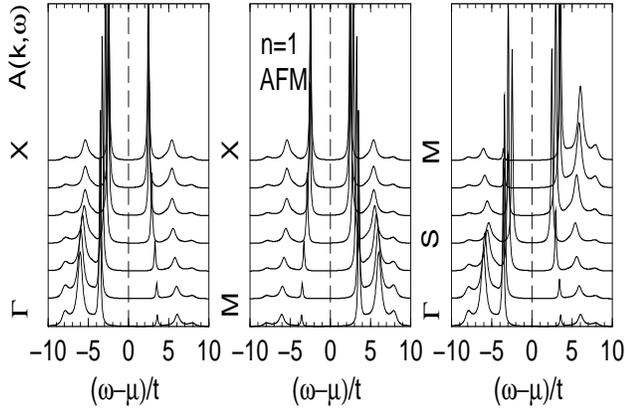,height=2.2in,width=3.3in}}
\narrowtext
\smallskip
\caption
{One-particle excitation spectra as obtained in the AF state at $n=1$ and 
 $T=0.05t$ for the Hubbard model with $U=8t$ ($t'=t''=0$).}
\label{fig:spectraaf}
\end{figure}
The spectral functions found within the DMFT (\ref{apes}) are dominated by
the lower Hubbard band (LHB), i.e., PES part at $\omega<\mu$, and the upper
Hubbard band (UHB), i.e., inverse photoemission (IPES) part at $\omega>\mu$,
separated by a large gap (Fig. \ref{fig:spectraaf}). Both
the PES and IPES spectrum show two distinct energy regimes:
 (i) narrow QP peaks at low energies, i.e., at the edge of the Mott-Hubbard
gap, and
(ii) incoherent and more extended features at higher energies $|\omega|>5t$.
The overall shape of the density of states $N(\omega)$ agrees very well with
the ED data for a $4\times 4$ cluster.\cite{Dag92}
The spectra have a characteristic ${\bf k}$-dependence with the overall
weight moving from the PES to IPES part along the $\Gamma-X-M$ and $\Gamma-M$
directions, where $\Gamma=(0,0)$, $X=(\pi,0)$ and $M=(\pi,\pi)$, in
qualitative agreement with QMC data.\cite{Sor88} The spectra obey the
particle-hole symmetry of the model, with spectra symmetric with respect
to $\omega=0$ at the $X$ and $S=(\pi/2,\pi/2)$ points. The spectrum at
the $M$ point is a mirror image of the one at the $\Gamma$ point.

\begin{figure}
\centerline{\psfig{figure=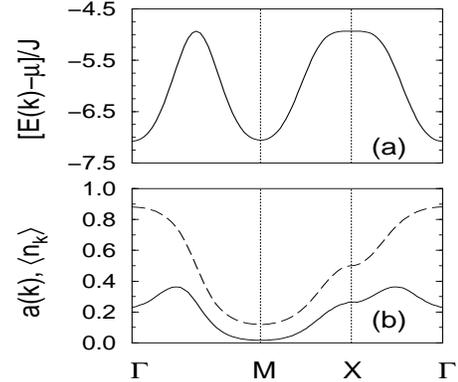,height=2.0in,width=2.3in}}
\narrowtext
\smallskip
\caption
{Momentum dependence in the main directions of the 2D BZ, as obtained for
 the PES spectrum of the
 Hubbard model at half-filling with $t'=t''=0$, $U=8t$, and $T=0.05t$:
 (a) QP dispersion $[E({\bf k})-\mu]/J$;
 (b) total electron occupation $\langle n_{\bf k}\rangle$ (dashed line)
     and QP weight $a({\bf k})$ (full line).}
\label{fig:band0}
\end{figure}

The QP maxima near the Mott-Hubbard gap resemble those found in the $t$-$J$
model in ED or within the self-consistent Born approximation,\cite{Dag94} in
spite of using a local self-energy in the present scheme. This shows that
the local many-body problem solved within the DMFT suffices to capture the
low-energy scale relevant for the QP propagation. Moreover, unlike in the
$t-J_z$ model which results in the ladder spectrum for a single
hole,\cite{Bul68,Mar91} the QP {\it can propagate\/}, as they couple to the
spin flips of the mean-field bath around site $i=0$ at which the many-body
problem is being solved. The QP dispersion is $\sim 2J$
[Fig. \ref{fig:band0}(a)], with the maxima along the AF Brillouin zone (BZ),
and remains very close to that found in the $t$-$J$ model.\cite{Dag94}

In the HF approximation, the electron occupation factors $\langle n_{\bf k}
\rangle$ are larger for the states which belong to the AF BZ than for the
remaining states outside the AF zone. On comparing the weights of the
electronic states with momenta ${\bf k}$ and ${\bf k}+{\bf Q}$, one finds
that also in the DMFT the electron weights are much larger within than
outside of the AF BZ [Fig. \ref{fig:band0}(b)]. The overall PES weight is
smoothly distributed in the 2D BZ, with the maximum (minimum) at the $\Gamma$
($M$) point, respectively. This result agrees well with a QMC simulations,
and the present data show the same step-like behavior of the electron
occupation factor $\langle n_{\bf k}\rangle$ when crossing the $X$ point
along the $\Gamma-X-M$ direction as the QMC data at $U=4t$ and
$8t$.\cite{Sor88,Bul94} 

\begin{table}[b]
\narrowtext
\caption{Values of the model parameters used for the presented calculations;
         the parameter sets chosen for La$_{2-x}$Sr$_x$CuO$_4$ and
         YBa$_2$Cu$_3$O$_{6+x}$ follow from the down-folding procedure of
         Ref. \protect\onlinecite{And95}.}
\vskip .5cm
\begin{tabular}[b]{ccccccc}
&       model parameters   & $t'/t$  & $t''/t$ &  $U/t$ &  $J/t$  & \\
\hline
& Hubbard model            &   0.0   &  0.0    &   8.0  &  0.50   & \\
& La$_{2-x}$Sr$_x$CuO$_4$  &  -0.11  &  0.04   &  10.0  &  0.40   & \\
& YBa$_2$Cu$_3$O$_{6+x}$   &  -0.28  &  0.18   &  12.0  &  0.33   & \\
\end{tabular}
\label{table1}
\end{table}
A similar step-like behavior is found as well in the QP weight $a_{\bf k}$ 
along the same line, determined by integrating the spectral functions 
(\ref{apes}) in an energy window of $2J$ which exhausts the range of the QP 
band in the density of states $N(\omega)$. 
The ${\bf k}$-dependence of the
QP weight is more complex than that of $\langle n_{\bf k}\rangle$ as two 
competing effects contribute along the $\Gamma-M$ and $\Gamma-X$ direction 
when the Mott-Hubbard gap is approached:
 (i) the QP pole moves to lower energies and thus the weight increases;
(ii) the overall PES weight is largest at the $\Gamma$ point and gradually
     decreases coming closer to the AF BZ.
Therefore, the maxima in the QP weight are found close to ${\bf k}=(\pi/2,0)$
and between the $\Gamma$ and $S=(\pi/2,\pi/2)$ point, while the (identical)
weights at the $X$ and $S$ point are lower. The lowest QP weight is found at
the $M$ point, but here instead a distinct QP exists in the IPES part, in 
agreement with the ED results.\cite{Esk96}

\begin{figure}
\centerline{\psfig{figure=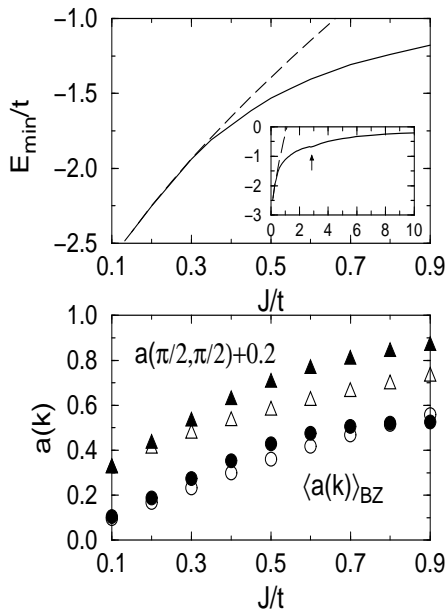,height=3.3in,width=2.3in}}
\narrowtext
\smallskip
\caption
{Quasiparticles in the AF state at $n=1$:
 the minimum of the QP band $E_{\rm min}/t$ (upper part), and the QP weight 
 $a({\bf k})$ at ${\bf k}=(\pi/2,\pi/2)$ and averaged over the BZ
 (lower part) as functions of $J/t$. Filled and empty symbols stand for 
 $a({\bf k})$ found in the present DMFT approach and in SCBA of Ref.
 \protect\onlinecite{Mar91}.
 The inset in the upper part shows $E_{min}/t$ for $0<J/t<10$; the value
 of $J/t$ at which the AF order vanishes is indicated by an arrow.}
\label{fig:qp}
\end{figure}
The QP weights $a({\bf k})$ increase with increasing $J/t$ and agree
surprisingly well with the self-consistent Born approximation and ED data 
for the $t$-$J$ model in the regime of $J/t<0.7$, as shown in Fig. 
\ref{fig:qp}. The average weight first increases somewhat faster than the 
numerical results of Ref. \onlinecite{Mar91}, but then flattens out above
$J/t\simeq 0.6$, and saturates indicating that the $t$-$J$ model does not
represent faithfully the hole dynamics in the Hubbard model at larger values
of $J/t$, where the excitations to the UHB become important. An equally good
agreement between the self-consistent Born approximation and ED data and the
present DMFT approach is found at individual ${\bf k}$-points; the values of
$a(\pi/2,\pi/2)$ are shown in Fig. \ref{fig:qp}, while a very good agreement
with ED data at $X$ point was presented earlier in Ref. \onlinecite{Fle98}.

The energy at the minimum of the polaron band found at the $S$-point follows
the power-law behavior found by Mart\'inez and Horsch \cite{Mar91} in the
range of $J/t<0.4$ (Fig. \ref{fig:qp}),
\begin{equation}
\frac{E_{\rm min}({\bf k}_S)}{t}=-3.20+2.94\left(\frac{J}{t}\right)^{0.702}.
\label{scaling}
\end{equation}
This power law supports the string picture, but is again closer to the full
single-hole problem in the $t$-$J$ model, where the data obtained from finite
cluster diagonalization could be fitted to the relation
$E_{\rm min}/t=-3.17+2.93(J/t)^{0.73}$ (see Ref. \onlinecite{Dag90}),
than to the $t-J_z$ model, which gives instead
$E_{\rm min}/t=-2\sqrt{3}+2.74(J_z/t)^{2/3}$ (see Ref. \onlinecite{Shr88}).
It is also quite close to the exact solution of the $t$-$J$ model in the
infinite-dimensional lattice, given by $E_{\rm min}/t=-4+2.34(J/t)^{2/3}$, 
which interpolates to the Nagaoka state.\cite{Str92}

\begin{figure}
\centerline{\psfig{figure=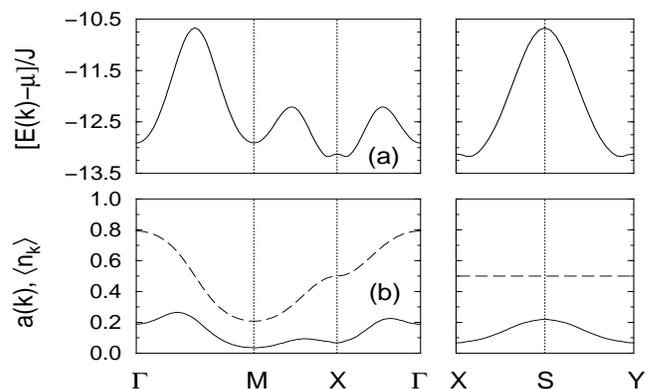,height=2.0in,width=3.3in}}
\narrowtext
\smallskip
\caption
{Momentum dependence in the main directions of the 2D BZ, as obtained for
 the PES spectrum of the Hubbard model at half-filling with extended hopping
 parameters $t'=-0.28t$, $t''=0.18t$, $U=12t$, and $T=0.05t$:
 (a) QP dispersion $[E({\bf k})-\mu]/J$;
 (b) total electron occupation $\langle n_{\bf k}\rangle$ (dashed line)
     and QP weight $a({\bf k})$ (full line).}
\label{fig:band1}
\end{figure}

Finally, we comment on the modifications of the spectra introduced by
the changes in the parameters $U$ and $t_{ij}$. Realistic parameters
for La$_{2-x}$Sr$_x$CuO$_4$ and YBa$_2$Cu$_3$O$_{6+x}$ were estimated both
using the cell method in the multiband charge-transfer model,\cite{Fei92}
and the down-folding procedure in the electronic structure
calculations.\cite{And95} 
Here we use the latter parameters as given in Table
\ref{table1}, but the sets do not differ significantly. By increasing the
value of $U$, one comes closer to the limit of the Heisenberg model, and
therefore the momentum density $\langle n_{\bf k}\rangle$ is more uniformly
distributed over the BZ [Fig. \ref{fig:band1}(b)]. This quantity depends
mainly on the ratio of $U/t$, and thus a similar result is obtained at the
same value of $U$ with $t^{\prime}$ and $t^{\prime\prime}$ non-zero.

In contrast, the earlier studies of the $t-t'-t''-J$ model have shown that 
the dispersion of QP's at low-energy are strongly dependent on the values of 
the extended hopping parameters, $t'$ and $t''$.\cite{Bal95}
This strong dependence is also found in the present calculations based on
the DMFT approach; the QP's at the $S$ and $X$ points are not degenerate
any more as soon as $t'\neq 0$. 
Here we present only the representative
result for larger values of $t'=-0.28t$ and $t''=0.18t$ found in
Sr$_2$CuO$_2$Cl$_2$, with minima located close to the $X$ point
[Fig. \ref{fig:band1}(a)]. 
Although the QP weight is dominated by the same
competition between the overall PES weight $\langle n_{\bf k}\rangle$ and the
position of the QP maximum with respect to the Fermi level, the consequences
of sizable $t'=-0.28t$ are clearly visible: the QP weight at the $X$-point is
reduced, and the degeneracy of the QP energies found before along the
$\Gamma-M$ and $X-\Gamma$ direction [Fig. \ref{fig:band0}(b)], respectively,
is now removed. As before, the lowest QP weight is found at the $M$ point,
and a distinct QP exists in the IPES part. Unlike at $t'=t''=0$, the latter
IPES spectrum is different from the PES spectrum at the $\Gamma$ point since 
there is no particle-hole symmetry at finite $t^{\prime}$.

\subsection{Spectral properties in spin-spiral states}
\label{sec:spectra}

As suggested by earlier 
studies,\cite{Shr89,Kub93,Zaa89,Arr91,Kan90,Fre92,Iga92} hole doping away 
from half-filling leads to incommensurate magnetic order. We found the same 
sequence of spiral phases with increasing doping as in the HF and slave-boson 
calculations:\cite{Arr91,Fre92} the AF order changes first into the SS with 
${\bf Q}=[\pi(1\pm 2\eta),\pi(1\pm 2\eta)]$ along the (1,1) direction [SS(1,1) 
state], and then at higher doping into the SS with ${\bf Q}=[\pi(1\pm 2\eta),
\pi]$ along the (1,0) direction [SS(1,0) state, or an equivalent SS(0,1) 
state]. The SS states with the components of the characteristic 
${\bf Q}$-vector shifted by $\pm 2\eta$ are physically equivalent and have 
the same energy. 
\begin{figure}
\centerline{\psfig{figure=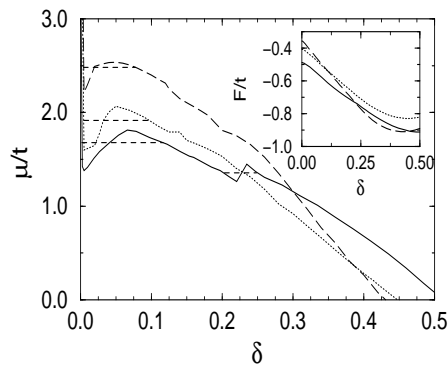,height=2.0in,width=2.3in}}
\narrowtext
\smallskip
\caption
{Chemical potential $\mu/t$ as a function of doping $\delta$, as obtained at
 $T=0.05t$ for three sets of parameters given in Table \protect\ref{table1}:
 the Hubbard model with $U=8t$ (full line) and for the model parameters of
 La$_{2-x}$Sr$_x$CuO$_4$ (dotted line), and YBa$_2$Cu$_3$O$_{6+x}$ 
 (long-dashed line).
 The regions of phase separation obtained from the Maxwell construction are
 indicated by dashed lines. The inset shows the free energy $F/t$ per site. }
\label{fig:mu}
\end{figure}
At fixed doping $\delta$ one finds, however, these phase 
transitions at larger values of $U$ in the present approach which includes 
local correlation effects, than in the effective single-particle 
theories.\cite{Arr91,Fre92} This change of the phase diagram follows from the 
correlation effects which screen the value of $U$ to $\bar{U}$ (\ref{ubar}), 
and strongly depend on the magnetic order (Table \ref{table2}). The highest 
value of effective $\bar{U}/U$
is obtained in the AF state at half-filling, where the double occupancy is
strongly reduced and the screening is thus ineffective. The screening is
stronger in the doped cases indicating that the moving electrons correlate
and avoid each other, leading to much weaker effective repulsion, and is
particularly pronounced in paramagnetic states. We found here a surprisingly
good agreement for the effective interaction $\bar{U}=1.98t$ found at $U=4t$
with the fitted value of $\bar{U}=2t$ in the QMC calculations.\cite{Bul93}

Two regions of phase separation which follow from the Maxwell
construction \cite{Arr91} were found for the Hubbard model at $U/t=8$
($t^{\prime}=t^{\prime\prime}=0$): a crossover regime from the AF to SS(1,1)
state for $0<\delta<0.11$, and from the SS(1,1) to SS(1,0) state for
$0.22<\delta<0.25$, respectively (Fig. \ref{fig:mu}). The value of the
chemical potential $\mu$ is $U/2$ at half-filling, and drops abruptly at
infinitesimal doping when it enters the LHB in a doped system. The doping
dependence of the free energy indicates a phase separation at low doping; 
this region becomes gradually narrower with increasing $U$, in agreement with
other calculations.\cite{Eme90,PvD95} In contrast, the transition to the
SS(1,0) state moves to larger doping with increasing $U$, and finally
disappears. Already at the model parameters of doped La$_2$CuO$_4$ we found
no region of stable SS(1,0) state. It is worth noting, however, that in this
case a small region of almost flat chemical potential $\mu$ was found for
$\delta\simeq 1/8$ which could be considered as a precursor effect for the
phase separation. It might lead to a different magnetic state at still lower
temperatures,\cite{Zaa89,Eme90} as the stripe structures observed in the
neutron experiments.\cite{Tra97}
\begin{table}[b]
\narrowtext
\caption{Values of magnetization $m_0$ (\protect\ref{moment}) and the
         renormalized interaction ${\bar U}$ (\protect{\ref{ubar}}),
         as obtained for the Hubbard model ($t'=t''=0$) at $T=0.05t$,
         $\delta=1-n$, for different magnetic states: antiferromagnetic (AF),
         spin spiral [SS(1,1) and SS(1,0)], and paramagnetic (PM) state.}
\vskip .5cm
\begin{tabular}[b]{ccccccc}
&  ground state & $\delta$ &  $U/t$ & $m_0$ & ${\bar{U}}/{U}$ \\
\hline
& AF      &  0.0    &  8  &  0.871  &  0.899  & \\
& AF      &  0.125  &  8  &  0.689  &  0.755  & \\
& SS(1,1) &  0.125  &  8  &  0.675  &  0.735  & \\
& SS(1,0) &  0.125  &  8  &  0.657  &  0.733  & \\
& AF      &  0.250  &  8  &  0.390  &  0.491  & \\
& SS(1,1) &  0.250  &  8  &  0.571  &  0.614  & \\
& SS(1,0) &  0.250  &  8  &  0.525  &  0.589  & \\
& PM      &  0.125  &  8  &  0.0    &  0.327  & \\
& PM      &  0.125  &  4  &  0.0    &  0.494  & \\
\end{tabular}
\label{table2}
\end{table}

\begin{figure}
\centerline{\psfig{figure=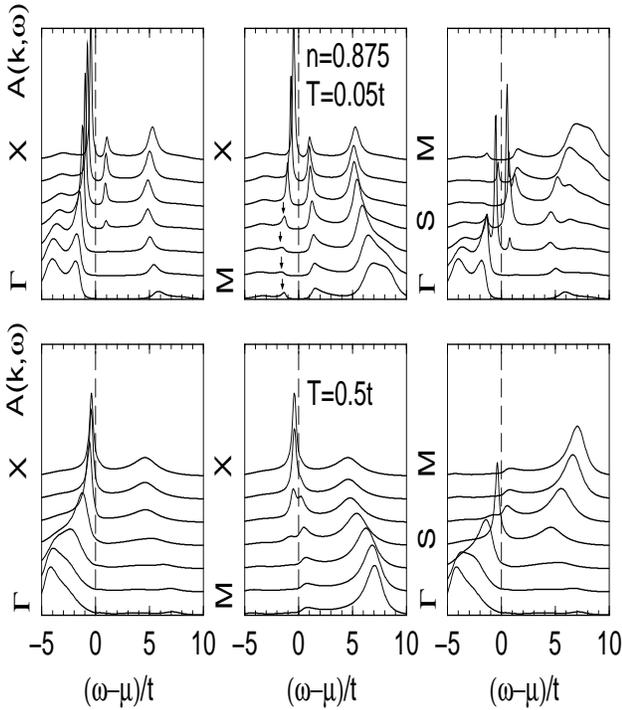,height=3.8in,width=3.3in}}
\narrowtext
\smallskip
\caption
{Spectral functions $A({\bf k},\omega)$ in the main BZ directions: $\Gamma-X$,
 $M-X$, and $\Gamma-M$ in SS(1,1) state at $\delta=0.125$ and $U=8t$ for:
 $T=0.05t$ (top) and $T=0.5t$ (bottom). The spectra along the $\Gamma-M$
 direction have been averaged over the (1,1) and ($\bar{1}$,$\bar{1}$)
 spirals, defined by ${\bf Q}=\pi(1-2\eta)(1,1)$ and 
 ${\bf Q}=\pi(1+2\eta)(1,1)$, respectively. A shadow band below $\mu$ in 
 $M-X$ direction at $T=0.05t$ is indicated by arrows.}
\label{fig:spectra8}
\end{figure}
\begin{figure}
\centerline{\psfig{figure=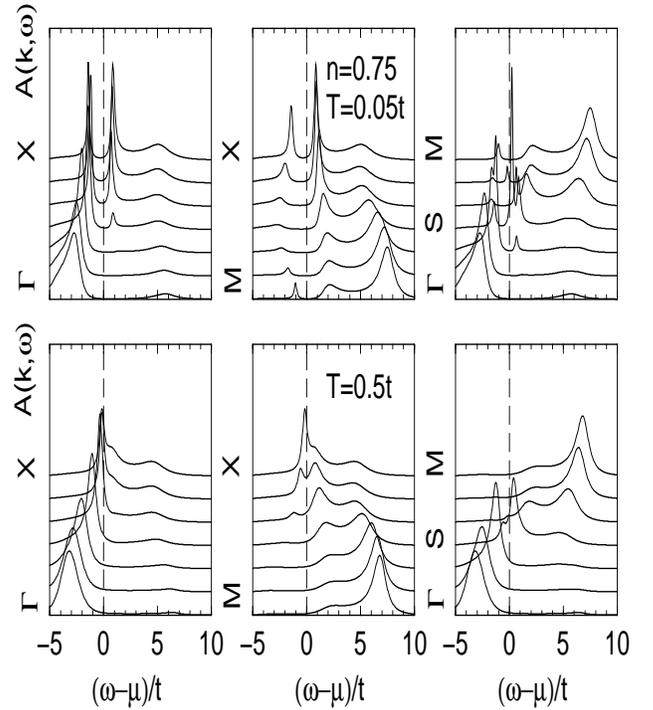,height=3.8in,width=3.3in}}
\narrowtext
\smallskip
\caption
{Spectral functions $A({\bf k},\omega)$ as in Fig.
 \protect{\ref{fig:spectra8}}, but for $\delta=0.25$ and (1,0) spiral at
 $T=0.05t$. The conventions are the same as in Fig.
 \protect{\ref{fig:spectra8}}. }
\label{fig:spectra4}
\end{figure}

The ${\bf k}$-resolved spectral functions (Figs. \ref{fig:spectra8} and
\ref{fig:spectra4}) allow us to identify the generic features of the doped
antiferromagnets described by the Hubbard model, in the regime of large $U$.
First of all, the spectra are still dominated by the large {\it Mott-Hubbard
gap\/} which separates the LHB from the UHB. The Mott-Hubbard gap develops
from the respective gap at half-filling and is considerably reduced from $U$
by the QP subbands which occur next to the large gap both in the LHB and in
the UHB. This large gap is accompanied by a smaller {\it pseudogap\/} $\sim
2t$ between the occupied (PES) and unoccupied (IPES) part of the LHB at low
temperature $T=0.05t$ (taking $t\simeq 0.4$ eV it corresponds to $\sim 200$
K). This pseudogap results from the SS order, and separates the majority and
minority spin states (with respect to the local coordinates at each site).
It is best visible along the $\Gamma-X$ and $X-M$ direction at $\delta=0.125$,
and becomes somewhat wider and less distinct in the SS(1,0) spiral at higher
doping $\delta=0.25$. We emphasize that the two features below and above the
chemical potential $\mu$ originate from the same QP peak at half-filling.
This shows that the QP found in the spectral function of one hole in the $t$-$J$
(or Hubbard) model cannot describe the regime with finite doping as the
{\it rigid band picture breaks down\/}.

The pseudogap is visible along the $\Gamma-X$ direction starting from
${\bf k}=(\pi/2,0)$, and the maximum above the chemical potential $\mu$ grows
gradually towards the $X$ point. Consider first the case of lower doping
$\delta=0.125$ (underdoped case). One finds that most of the
spectral weight at the $X$ point is still at $\omega<\mu$, with a sharp QP
peak at $\omega\simeq -0.44t$. Increasing ${\bf k}$ along the $X-M$ direction
gives a transfer of the overall weight to higher energies, and the QP peak
below $\mu$ gradually looses the intensity, while the peak above $\mu$ takes
over around ${\bf k}=(\pi,\pi/3)$. However, the feature at $\omega<\mu$ is
still well visible as a 'shadow' QP band (Fig. \ref{fig:spectra8}), with
a width $\sim 2J$. Thus, the QP band of the $t$-$J$ model is drastically
modified at finite doping, and a new energy scale $\sim 1.5t$ due to the
pseudogap accompanies the dispersive QP feature below the chemical potential.

A similar situation is found also at higher doping $\delta=0.25$ (overdoped
regime), and the pseudogap is quite pronounced along the $\Gamma-X$ and $X-M$
directions (Fig. \ref{fig:spectra4}). However, except for the neighborhood of
the $\Gamma-$ point, more spectral weight is found at high energies. Already
at the $X$-point one finds that the peak at $\omega\simeq t$ has a higher
intensity than the one below $\mu$. It becomes gradually weaker when the
$M$ point is approched, and disperses in the energy range $\sim 3J$, while
the feature below $\mu$ still has a similar dispersion $\sim 2J$ as in the
$\delta=0.125$ case. We note that the pseudogap increases to $\sim 2.5t$.
Moreover, one finds that the QP dispersion is broader at $\delta=0.25$,
indicating the gradual weakening of the local magnetic order with increasing
doping.

The spectra are drastically changed, in particular in the low-energy range
of $|\omega-\mu|<2t$ when the temperature is increased. At $T\simeq 0.3t$
the SS order is unstable against the AF order which we interpret as a
crossover to the small regions of the short-range order with the preferably
AF ordering of nearest-neighbor spins. The spectra found for doping
$\delta=0.125$ at $T=0.5t$ consist of broad maxima which correspond to the
LHB and UHB, respectively, and only a single maximum is found in
$A({\bf k},\omega)$ next to the $X$-point. These data, and also the spectral
functions for $T=0.33t$ reported earlier,\cite{Fle98} agree remarkably well
with the results of QMC calculations.\cite{Pre97} 
The spectra at $\delta=0.25$ and $T=0.5t$ are quite similar to those at
lower doping $\delta=0.125$, with more weight in the IPES part of the LHB.

\begin{figure}
\centerline{\psfig{figure=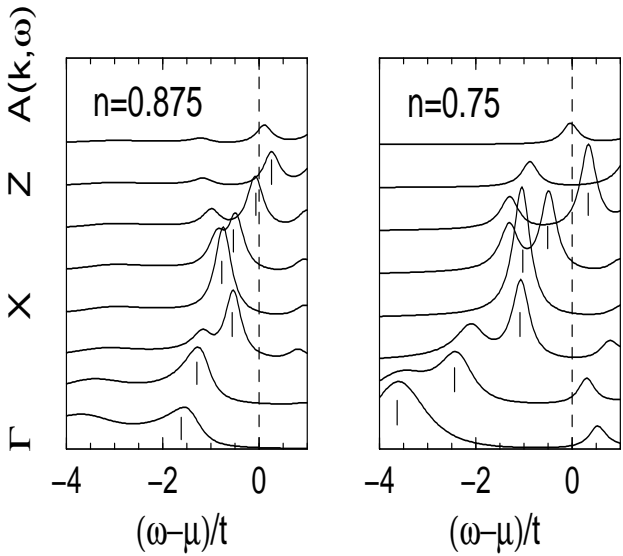,height=3.0in,width=3.3in}}
\narrowtext
\smallskip
\caption
{Spectral functions $A({\bf k},\omega)$ along two main directions in a 2D BZ:
 $\Gamma-X$ [with the step of $(\pi/3,0)$], and $X-Z$, where $Z=(\pi,\pi/2)$
 [with the step of $(0,\pi/6)$], in SS(1,1) state as obtained for the model
 parameters of doped La$_{2-x}$Sr$_x$CuO$_4$ (Table \protect\ref{table1}) at
 doping $\delta=0.125$ (left) and $\delta=0.25$ (right), and after averaging
 over all equivalent SS states with different values of ${\bf Q}$. The 
 dispersive feature with the strongest intensity is indicated by vertical 
 lines.}
\label{fig:lsco}
\end{figure}
\begin{figure}
\centerline{\psfig{figure=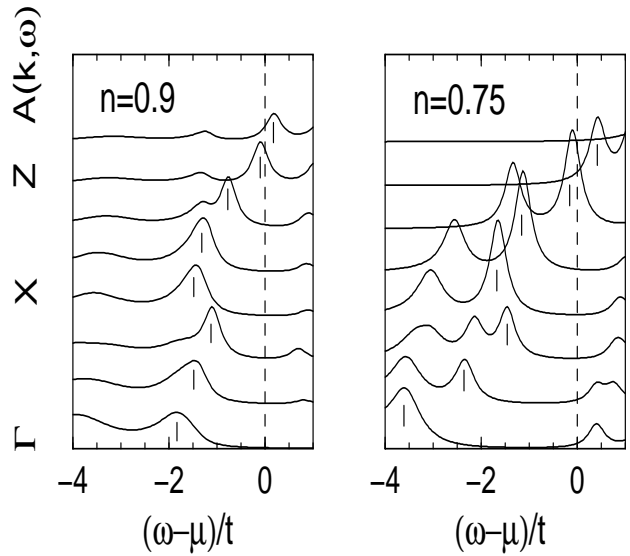,height=3.0in,width=3.3in}}
\narrowtext
\smallskip
\caption
{The same as in Fig. \protect\ref{fig:lsco}, but for the model
 parameters of doped Y(Bi)-superconductors (Table \protect\ref{table1}) at
 doping $\delta=0.10$ (left) and $\delta=0.25$ (right), and after averaging
 over all equivalent SS states with different values of ${\bf Q}$.}
\label{fig:bisco}
\end{figure}

We do not intend to present detailed analysis of the spectra obtained using
the extended hopping parameters which correspond to the electronic structure
of La$_{2-x}$Sr$_x$CuO$_4$ and YBa$_2$Cu$_3$O$_{6+x}$, respectively. Instead,
we point out the important similarities and differences to the Hubbard model
as far as the SS states are concerned. Consider first the effective
parameters of La$_{2-x}$Sr$_x$CuO$_4$. First of all, a narrow QP band is
also found below the Fermi energy (Fig. \ref{fig:lsco}), but the measured
dispersion between the $\Gamma$ and $X$ point is $\sim 0.80t$ ($\sim 2.56t$)
at $\delta=0.125$ ($\delta=0.25$), while it amounts only to $\sim 0.36t$
at $\delta=0$. Note that the energies of the QP peak are much closer to each
other for the present parameters than in the Hubbard model with $t'=t''=0$,
where one finds instead the dispersion of $1.37t$ ($3.58t$) at $\delta=0.125$
($\delta=0.25$) at $U=8t$, while it amounts to $1.1t$ in the undoped case.
This gradual widening of the QP dispersion with increasing doping may be 
understood as a consequence of the admixture of ferromagnetic components with 
increasing doping in the SS(1,1) states. The same trend is also observed for 
the parameters of YBa$_2$Cu$_3$O$_{6+x}$, where one finds the QP states in 
the PES part separated by $\sim 0.39t$ ($\sim 1.94t$) between the
$\Gamma$ and $X$ point at $\delta=0.10$ (0.25) (Fig. \ref{fig:bisco}), while
this splitting is only $\sim 0.07t$ at half-filling.

Finally, the finite hopping elements to the second and third neighbors
stabilize the SS(1,1) state with respect to the SS(1,0) state also at higher
doping (see Fig. \ref{fig:mu}), and therefore the intensity at the $X$ point
does not cross the Fermi level even at $\delta=0.25$ for both parameter sets.
In fact, taking $J=0.125$ meV ($t/J=3$), the QP state at the $X$ point is
found at $\omega\simeq -0.56$ eV, and does not change significantly as a
function of doping (Fig. \ref{fig:bisco}). In contrast, in the ARPES
experiments for Bi$_2$Sr$_2$CaCu$_2$ the QP state at $X$-point is found at
energy $\simeq -0.20$ eV ($\simeq -0.056$ eV) in the underdoped (optimally
doped) compound.\cite{Mar96} This indicates that either an improved solution
of the many-body problem is still required, or the actual magnetic order
in these compounds might be different from SS states. However, the observed
increase of the onset of incommensurability with increasing $U$ and
$t^{\prime}$ is consistent with the observations made by Igarashi and Fulde
\cite{Iga92} and with QMC calculations of Duffy and Moreo.\cite{Duf95}

\subsection{Total densities of states}
\label{sec:dos}

\begin{figure}[t]
\centerline{\psfig{figure=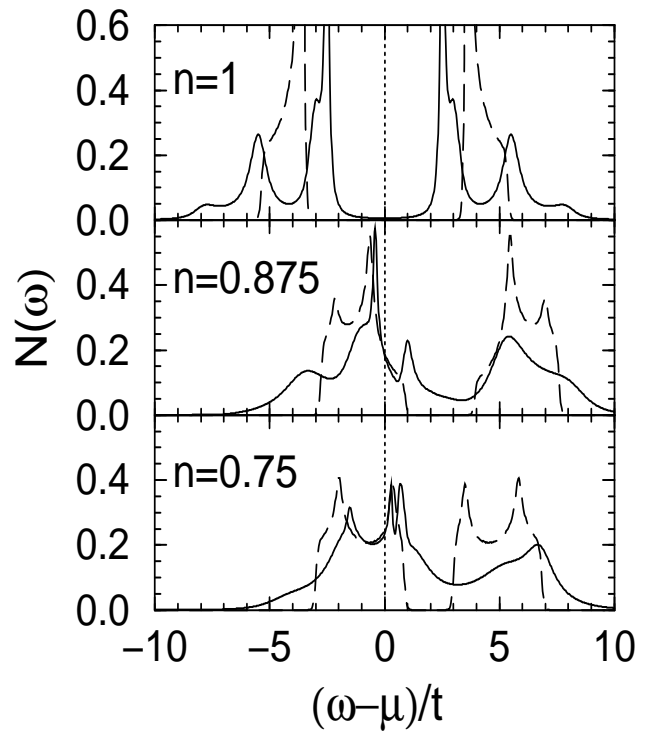,height=4.0in,width=3.3in}}
\narrowtext
\smallskip
\caption
{Total densities of states $N(\omega)$ as obtained within DMFT (full
 lines) for $\delta=0$ (AF state), 0.125 [(1,1) spiral], and 0.25 [(1,0)
 spiral] with $U/t=8$ and $T=0.05t$. The dashed lines show $N(\omega)$
 for the magnetic ground states found in the HF approximation. }
\label{fig:dos}
\end{figure}

We already pointed out\cite{Fle98} a very good agreement between the
calculated density of states (\ref{dos}) and the results of ED by Dagotto
{\it et al\/}.\cite{Dag92} Here we present instead a comparison between the
density of states obtained within the DMFT method and that found in the HF
approximation (Fig. \ref{fig:dos}). First of all, one notices a narrower gap
which separates the QP subbands in the DMFT of width $\sim 2J$, instead of
the HF one-particle states, on the scale of $\sim 2t$. This part of the
spectral density might also be reproduced in effective single-particle
approaches, as for instance in the slave-boson mean-field theory. However,
the incoherent parts which extend on the energy scale down to
$|\omega-\mu|\simeq 9t$ result from many-body scattering and can only be
reproduced if the dynamical part of the self-energy is included. The overall
width of the subbands at $\omega<\mu$ and $\omega>\mu$ is $\sim 7t$,
respectively, as known from the analysis of the $t$-$J$ model in ED and in QMC
calculations.\cite{Dag94}

It is evident that due to the changes of $N(\omega)$ in the range of
$|\omega-\mu|\leq 1.5t$ with respect to QP band in the undoped system, the
low-energy part of the spectrum cannot be reproduced in a renormalized
one-particle theory. The pseudogap in the doped systems is not visible in
the HF density of states, and it remains a challenge whether an effective
one-particle theory which captures this essential new energy scale could be
constructed. As expected, the agreement between the HF and DMFT density of
states improves somewhat at higher doping $\delta=0.25$, where the
Mott-Hubbard gap is gradually lost, and the system approaches the
single-particle limit. We note, however, that the gap between the LHB and
the UHB relies in our approach on the magnetic order, and a more accurate
approach in the strongly doped regime at large $U$ would instead have to 
include the scattering on local moments.

\begin{figure}
\centerline{\psfig{figure=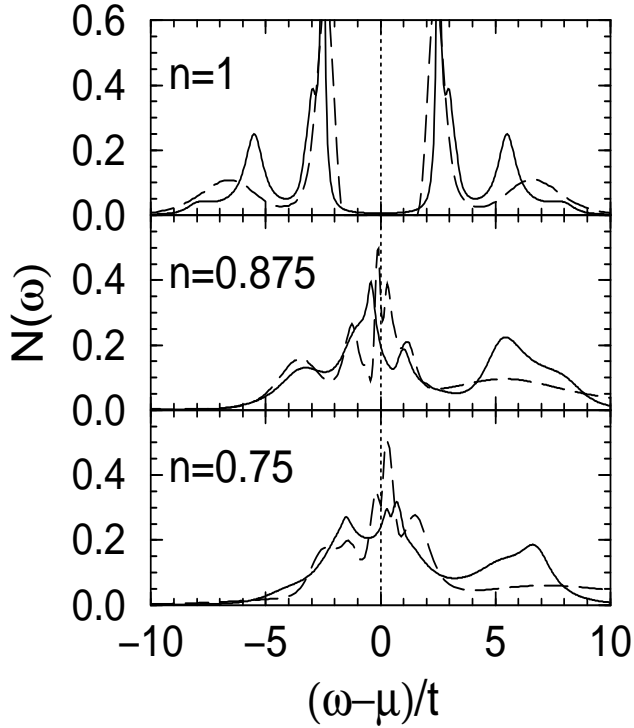,height=4.0in,width=3.3in}}
\narrowtext
\smallskip
\caption
{Total densities of states $N(\omega)$ as obtained within DMFT at $U=8t$ by
 calculating the self-energy either using an analytic formula
 (\protect\ref{sigma}) (solid lines), or by DMFT-QMC method (dashed
 lines). Different panels show the results obtained for $\delta=0$ (AF state),
 0.125 [(1,1) spiral], and 0.25 [(1,0) spiral], respectively, at $T=0.05t$. }
\label{fig:qmc}
\end{figure}

In spite of a very good agreement between the present DMFT approach and the
ED data,\cite{Fle98} it is interesting to investigate to what extent the
analytic formula for the self-energy (\ref{sigma}) describes the hole
dynamics in a doped system. Therefore, we performed also a DMFT-QMC
calculation of local $\hat{\Sigma}_{\bf Q}(i\omega)$ for SS states, and the 
corresponding densities of states, shown in Fig. \ref{fig:qmc}. The QP peaks 
are very close to each other at half-filling, while the incoherent states at 
higher energies in the LHB and UHB have almost the same weights, but are 
moved to somewhat higher energies in the QMC calculation. The increase of the 
spectral weight close to the Fermi level is well pronounced in the latter 
calculation at $\delta=0.125$, but one finds instead {\it a pseudogap smaller 
by a factor close to five\/}. 
However, one should realize that the present
calculation performed at low temperature $T=0.05t$ corresponds in practice to
the ground state, while the same temperature in QMC includes already thermal
fluctuations which considerably reduce the size of the pseudogap. Indeed, we
find using the ED method to solve the self-energy within DMFT the pseudogap
in the SS state of $\sim 0.7t$. It might be expected that this reduction of 
the energy scale would result in a better quantitative description of the 
spectral functions and the related excitations across the pseudogap, leading 
to a reduced energy scale for the low-energy features of the optical 
conductivity (Sec. \ref{sec:optic}). We also found more extended energy range 
of the incoherent states which belong to the UHB in the QMC calculation.
Altogether, the comparison with the DMFT-QMC calculation demonstrates that
the analytic method developed in this paper is very useful for a qualitative
insight into the possible changes of magnetic states under doping and their
consequences for the properties measured in experiment.

\section{Optical and transport properties}
\label{sec:optic}

\subsection{Optical conductivity in the Hubbard model}
\label{sec:hubbard}

The evolution of the spectral functions $A({\bf k},\omega)$ and the density
of states $N(\omega)$ with doping, reported in Secs. \ref{sec:mott} and
\ref{sec:pes}, motivates an investigation of the optical properties. Here we
make use of the theory introduced in Sec. \ref{sec:transport}, where we have
shown how the optical conductivity can be derived from the local self-energy
in the present DMFT treatment.

\begin{figure}
\centerline{\psfig{figure=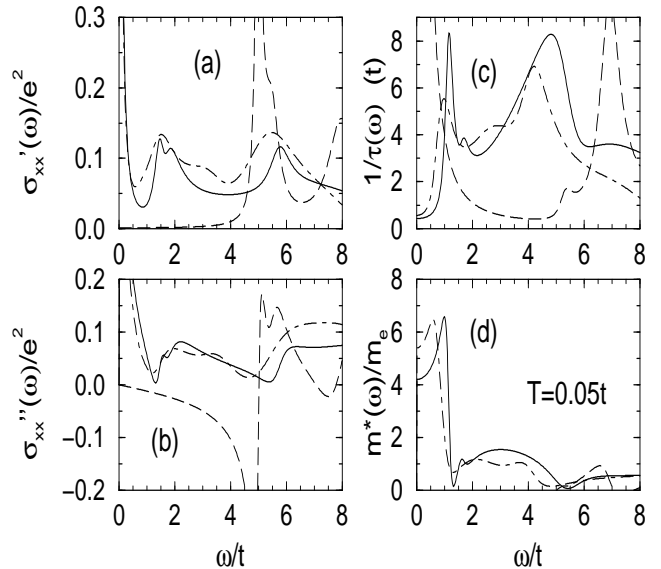,height=3.1in,width=3.3in}}
\narrowtext
\smallskip
\caption
{Optical properties as functions of energy $\omega/t$ for the Hubbard model
 with $U=8t$ at low temperature $T=0.05t$ for
 $\delta=0$     (dashed lines),
 $\delta=0.125$ (  full lines), and
 $\delta=0.25$  (dashed-dotted lines):
 (a) real      part of the optical conductivity $\sigma' (\omega)$;
 (b) imaginary part of the optical conductivity $\sigma''(\omega)$;
 (c) scattering rate $1/\tau(\omega)$;
 (d) effective mass $m^*(\omega)/m_e$. }
\label{fig:opticlo}
\end{figure}

\begin{figure}
\centerline{\psfig{figure=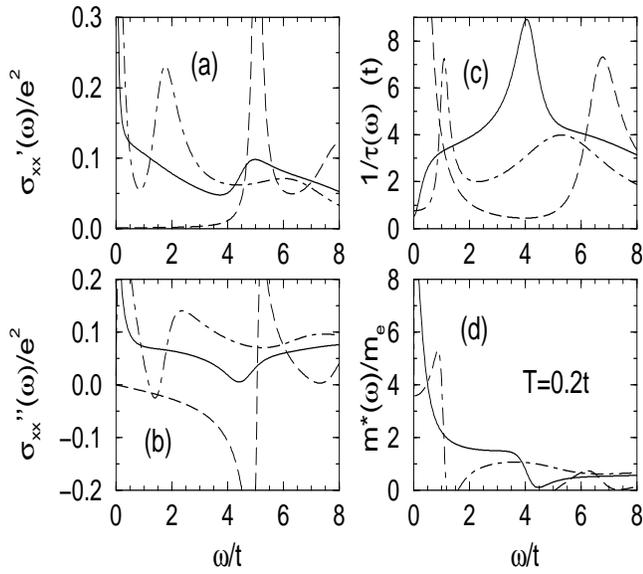,height=3.1in,width=3.3in}}
\narrowtext
\smallskip
\caption
{Optical properties as functions of energy $\omega/t$ for the Hubbard model
 with $U=8t$ at intermediate temperature $T=0.2t$;
 the meaning of lines as in Fig. \protect{\ref{fig:opticlo}}.}
\label{fig:optichi}
\end{figure}

As an illustrative example, we concentrate on the optical
conductivity found for the Hubbard model with the nearest-neighbor hopping
($t^{\prime}=t^{\prime\prime}=0$) and $U=8t$. We present the optical data in
Figs. \ref{fig:opticlo} and \ref{fig:optichi} at two temperatures: $T=0.05t$ 
and $0.2t$. While the magnetic order is AF at half-filling, the SS states 
characterized by the wave-vector 
${\bf Q}=[\pi(1\pm 2\eta_x),\pi(1\pm 2\eta_y)]$ change with doping and 
temperature. At lower doping $\delta=0.125$ we find a SS(1,1) with 
$\eta_x=\eta_y=0.125$ (0.09) at $T=0.05t$ ($T=0.2t$), respectively, while at 
higher doping $\delta=0.25$, a SS(1,0) state ($\eta_y=0$) with $\eta_x=0.25$ 
(0.23), or an equivalent SS(0,1) state, is found instead.

At half-filling one finds a large gap below $\omega\simeq 4.9t$ at $U/t=8$
and no Drude peak which shows that the system is in the insulating
phase.\cite{Sca93} The conductivity at $\omega>4.9t$ is incoherent and
originates from the excitations across the Mott-Hubbard gap.
This changes drastically when the system is doped and two new features
occur at lower energy: the Drude peak, and the mid-gap state at
$\omega\simeq 2t$, both with the increasing intensity between
$\delta=0.125$ and $\delta=0.25$ at low temperature (Fig. \ref{fig:opticlo}).
These features are accompanied by an incoherent background of the excitations
within the LHB. The peak at $\omega\simeq 2t$ corresponds to 
excitations across the pseudogap; as such it is more influenced by the
increasing temperature in the underdoped regime, where the SS(1,1) state
is less robust than the SS(1,0) state in the overdoped regime.

Below $\omega=4.9t$ the frequency dependent scattering rate $1/\tau(\omega)$
and the effective mass $m^*(\omega)/m_e$ can also be divided into two regions:
 (i) above $\omega\simeq 2t$ the scattering rate increases monotonically with
increasing frequency,
(ii) below $\omega\simeq 2t$ it has a maximum at energy 
$\omega\simeq 1.15t$ ($1.0t$) for
$\delta=0.125$ ($0.25$), and drops to zero for $\omega\to 0$ at finite
doping. This behavior for $\omega\to 0$ and $T\to 0$ is consistent with the
Fermi liquid behavior which follows from the local approximation to the
self-energy (\ref{localg}). A finite value at $\omega=0$ is a numerical
effect due to finite broadening of the spectra ($\epsilon=0.1t$).

The frequency region in which the scattering is suppressed has a direct
relation to the existence of a pseudogap region in the single-particle 
spectral function $A({\bf k},\omega)$, reported in Sec. \ref{sec:spectra}, 
and indicates that SS LRO reduces the scattering of the charged carriers in 
the energy range $\omega<1.15t$. At the same time, the effective mass
$m^*(\omega)$ rises to a maximum value of $\sim 5m_e$ within the pseudogap
region, and is found to be rather independent of hole doping. As the 
temperature increases to $T=0.2t$, the pseudogap disappears and the region 
of suppressed scattering is filled up in the underdoped regime with 
$\delta=0.125$, while the low scattering persists for $\omega<1.0t$ at 
$\delta=0.25$ (see Fig. \ref{fig:optichi}). At the same time, the mid-gap 
state in the real part of the optical conductivity changes into a smooth 
feature which extends down to the Drude peak for $\delta=0.125$, contrary 
to the case with $\delta=0.25$ where the spectral weights of the above two 
features remain well separated. This is clearly related to the behavior 
observed in $A({\bf k},\omega)$ with increasing temperature, where the 
pseudogap along the $X-M$ direction filled up with spectral weight as $T$ 
increased for $\delta=0.125$. 

These changes of the mid-gap state with temperature are due to the changes of 
the magnetic correlations in the doped systems included in our calculations
which do not distinguish between long-range and short-range magnetic order,
but treat local dynamical correlations. However, there are indications that 
the mid-gap feature results from an interplay between short-range magnetic 
order and electron correlations.\cite{Cas92} Therefore, the rather 
strong evolution of the low-energy weight with increasing temperature shown 
in Figs. \ref{fig:opticlo} and \ref{fig:optichi} may be overestimated in the 
present treatment of the self-energy which does not allow to get a 
metal-insulator transition {\em without\/} an accompanying magnetic LRO. 
We also note that the mid-gap states are likely a bare consequence of the 
strongly correlated nature of optical and one-particle excitations in the 
Hubbard model,\cite{Cas92} and it is still a challenge to describe them 
better in a theory which would treat the AF and paramagnetic states with
local moments on equal footing.

The frequency-dependent scattering rate allows us to find a {\it
crossover temperature\/} $T^*$ at which the pseudogap closes. We estimated
that $T^*\simeq 0.26t$ for $\delta=0.125$, and observed a monotonic increase
of $1/\tau(\omega,T^*)$ up to $\omega\sim 4.1t$. At $T=0.2t$ the effective
mass increases up to $\sim 10m_e$ within the pseudogap at $\delta=0.125$.
At half-filling and $T=0.2t$ one finds that the charge-transfer gap is only
slightly reduced from its value at $T=0.05t$, and the insulating behavior is
accompanied by AF LRO. We estimated the N\'eel temperature for $U=8t$ to
be $T_N\simeq 0.62J$.

Further evidence for a characteristic crossover temperature $T^*$ may be
found in the behavior of the in-plane dc resistivity (\ref{dcr}).
The resistivity received a lot of attention in connection with the observed
normal state pseudogap in the electronic excitation spectrum,\cite{Bat94}
and from theoretical point of view.\cite{Jak94,Jar94} In fact,
the physical origin of linear $T$-dependence of $\rho(T)$ for samples of
high-$T_c$ compounds close to the optimal doping level remains puzzling.

\begin{figure}
\centerline{\psfig{figure=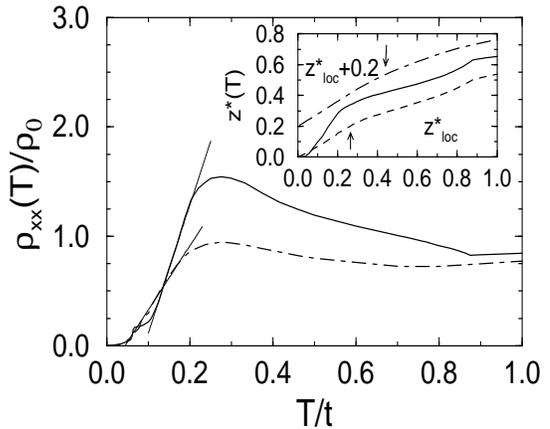,height=2.4in,width=2.8in}}
\narrowtext
\smallskip
\caption
{Resistivity $\rho(T)/\rho_0$ as a function of temperature $T/t$ as
 obtained for the Hubbard model with $U=8t$ for
 $\delta=0.125$ (full line) and $\delta=0.25$ (dashed line).
 The inset shows the weight $z^*(T)$ (\protect{\ref{zstar}}) found at the
 $X$ point at $\delta=0.125$ (full line), and averaged weights over the
 Brillouin zone at $\delta=0.125$ (dashed line) and at $\delta=0.25$
 (dashed-dotted line). Arrows in the inset indicate $T^*$.}
\label{fig:zstar}
\end{figure}
The results for $\rho_{xx}(T)$ obtained for the Hubbard model at two doping
levels, $\delta=0.125$ and $\delta=0.25$, are shown in Fig. \ref{fig:zstar}.
At low temperatures, $T<0.06t$, the resistivity shows Fermi-liquid behavior
for both hole densities, i.e., $\rho_{xx}(T)\propto T^2$. As usually in the
DMFT calculations,\cite{Jar94} the $T^2$-dependence of $\rho_{xx}(T)$
originates from the low-frequency behavior of the imaginary part of the local
self-energy. In the regime of high temperatures ($T > 0.9t$), the resistivity 
increases linearly with temperature which is due to temperature independence 
of the spectral functions $A_{\sigma\sigma^\prime}({\bf k},\omega)$ at
$T\to\infty$, and the high temperature limit of the derivative,
$\left( -\partial n_{\rm F}(\omega)/\partial\omega\right)\to 1/(4T)$, thus
leading to $\sigma_{xx}(T)\propto 1/T$, i.e., $\rho_{xx}(T)\propto T$. As the
temperature is lowered, the magnetic moments gradually build up, and a kink
in the resistivity appears. Therefore, the increase in the resistivity as the
temperature is lowered can be attributed to the enhancement in the scattering
of electrons by local spin fluctuations.

On the contrary, for large hole doping the system is a better metal, hole
spin correlations are gradually lost, and the increase of resistivity is less
pronounced in the temperature region $T<0.88t$ at $\delta=0.25$.
The maximum of $\rho_{xx}(T)$ for $\delta=0.125$ is located almost exactly at
$T\approx 0.26t$ ($\sim 750 K$ taking the experimental value of the
superexchange $J=125$ meV), where the pseudogap in the single-particle
excitation spectra opens leading to a suppression of the effective scattering
rate $1/\tau(\omega,T)$, as discussed previously. This defines the crossover 
temperature $T^*$. Remarkably, the change from a linear to a nearly-linear 
$T$-dependence, $\rho_{ab}(T)\propto T^{1+\epsilon}$ ($\epsilon>0$), of the 
in-plane dc resistivity of La$_{2-x}$Sr$_x$CuO$_4$ was found to be at 
$T^*\simeq 600$ K for $x\simeq 0.13$ and was attributed to the opening of a 
pseudogap in the electronic excitation spectrum.\cite{Bat94} However, the 
saturation of the resistivity $\rho_{ab}(T)$ cannot be observed in a real 
system as the carriers also couple to other bosonic excitations, e.g. to 
phonons, which are neglected here.

Upon lowering the temperature below $T^*$, at $T<0.24t$ for both
$\delta=0.125$ and for $\delta=0.25$, we observed a nearly linear
$T$-dependence of $\rho_{xx}(T)$. In this temperature range the SS wave
vector, ${\bf Q}=[\pi(1\pm 2\eta_x),\pi(1\pm 2\eta_y)]$, becomes strongly
temperature dependent, and maintains the directional deviation from the AF
wave vector, ${\bf Q}_{\rm AF}=(\pi,\pi)$, with $\eta_x=\eta_y=\eta(T)$ for
$\delta=0.125$ and $\eta_x=\eta(T)$ ($\eta_y=0$) for $\delta=0.25$. 
In both cases $\eta(T)$ 
increases from $\eta(T^*)\simeq 0$ with decreasing temperature and saturates 
at its ground-state value $\eta(T=0)\simeq \delta$ below $T\simeq 0.08t$. In 
the linear regime ($T<T^*$) the resistivity can be fitted quite well by a 
linear $T$-dependence, as expected for the SS states,\cite{Shr89}
$\rho_{xx}^{\rm fit}(T)=\rho_{xx}^{\rm fit}(0)+\zeta\rho_0\delta^{-1}T$,
with $\rho_{xx}^{\rm fit}(0)/\rho_0=-1.05$ (-0.25) for $\delta=0.125$ (0.25),
respectively, where the increase of the negative temperature coefficient
$\rho_{xx}^{\rm fit}(0)$ is a further manifestation of the gradual loss of
local magnetic moments as doping is increased. On the contrary, in the
paramagnetic phase of the Hubbard model at $d=\infty$ one finds
$\rho_{xx}^{\rm fit}(0)\geq 0$.\cite{Jar94} Furthermore, the slope of
$\rho_{xx}(T)$ in the low temperature regime is given by $\zeta\simeq 1.46$
independent of hole density. This value is larger by about a factor of 2.5
than the respective slope found in the retraceable-path approximation, 
\cite{Bri70} and in the ED studies at finite temperature,\cite{Jak94} being 
$\zeta=0.55$ and 0.60, respectively, and demonstrates that the changes in 
the magnetic order with increasing temperature influence significantly the 
system resistivity. Unfortunately, such effects cannot be studied in the ED 
method due to the small size of considered clusters.

In order to further support our observation that the crossover temperature
$T^*$ is related to the pseudogap in the single-particle excitation spectrum
we plot in the inset of Fig. \ref{fig:zstar} an average of the single-particle
spectral weight within an energy window $\propto T$ around the Fermi energy
$\omega=0$, defined by,\cite{Tri95,Vil97}
\begin{eqnarray}
z^*(T)&=&-\sum_{\sigma \sigma^\prime}
 G_{\sigma \sigma^\prime {\bf Q}}({\bf k}_{X},\tau=\beta/2)     \nonumber \\
 &=& \frac{1}{2} \int_{-\infty}^{\infty} d\omega \;
 \frac{A({\bf k}_{\rm F},\omega)}{\cosh(\beta \omega/2)}.
\label{zstar}
\end{eqnarray}
Similarly, a measure for the temperature dependence of the density of states
at the Fermi energy $N(0)$ is obtained from the local Green's function
(\ref{localg}),
$z^*_{loc}(T)=-\sum_{\sigma}G_{\sigma\sigma {\bf Q}}(\tau=\beta/2)$.
In the low temperature limit $N(0)$ can be obtained from the relation
$N(0)\simeq\beta z^*_{loc}(T)/\pi$,\cite{Tri95} which gives $\simeq 0.20$
($\simeq 0.26$) for $\delta=0.125$ ($\delta=0.25$), respectively. However,
one finds that the one-particle density of states at the Fermi energy does
not evolve smoothly to the low temperature values, but instead states are
depleted from the region $\omega\simeq\mu$ as $T$ is reduced below
$T^*\simeq 0.26t$ ($0.43t$) for $\delta=0.125$ ($\delta=0.25$), respectively.
In particular, we observed a faster loss of the QP weight with
momentum ${\bf k}_{X}=(\pi,0)$ for $\delta=0.125$ (Fig. \ref{fig:zstar}).
This shows that the opening of the pseudogap in the one-particle excitation
spectrum at $(\pi,0)$ coincides with the suppression of the effective
scattering rate $1/\tau(\omega,T)$.

Experimentally, the resistivity changes from a linear to a nearly-linear 
$T$-dependence at $T^*$ of the order of 500 K. Although our calculations do 
not allow to interpret the linear part of $\rho_{xx}(T)$ at high temperature 
$T>T^*$ as only the electronic degrees of freedom are included, 
we note that the enhanced slope of $\rho_{xx}(T)$ at low temperature 
$T\sim 100$ K and the negative temperature coefficient agree qualitatively
with the experimental results for YBa$_2$Cu$_3$O$_{7-x}$ in the underdoped
regime.\cite{Ito93} Our calculations confirm the conjecture of Shraiman and
Siggia of a nearly-linear $T$-dependence of the resistivity for a system with 
SS magnetic order.\cite{Shr88} These features can be seen as generic 
fingerprints of incommensurate magnetic correlations.

\subsection{Implications of extended hopping}
\label{sec:tprime}

\begin{figure}
\centerline{\psfig{figure=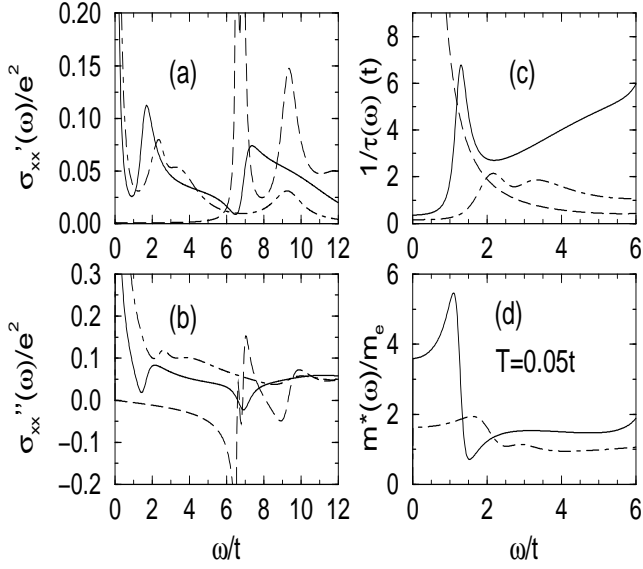,height=3.1in,width=3.3in}}
\narrowtext
\smallskip
\caption
{Optical properties as functions of energy $\omega/t$ for the Hubbard model
 with extended hopping parameters of La$_{2-x}$Sr$_x$CuO$_4$ (Table
 \protect\ref{table1}) at low temperature $T=0.05t$ for
 $\delta=0$     (dashed lines),
 $\delta=0.125$ (  full lines), and
 $\delta=0.25$  (dashed-dotted lines):
 (a) real      part of the optical conductivity $\sigma' (\omega)$;
 (b) imaginary part of the optical conductivity $\sigma''(\omega)$;
 (c) scattering rate $1/\tau(\omega)$;
 (d) effective mass $m^*(\omega)/m_e$. }
\label{fig:opticla}
\end{figure}
\begin{figure}
\centerline{\psfig{figure=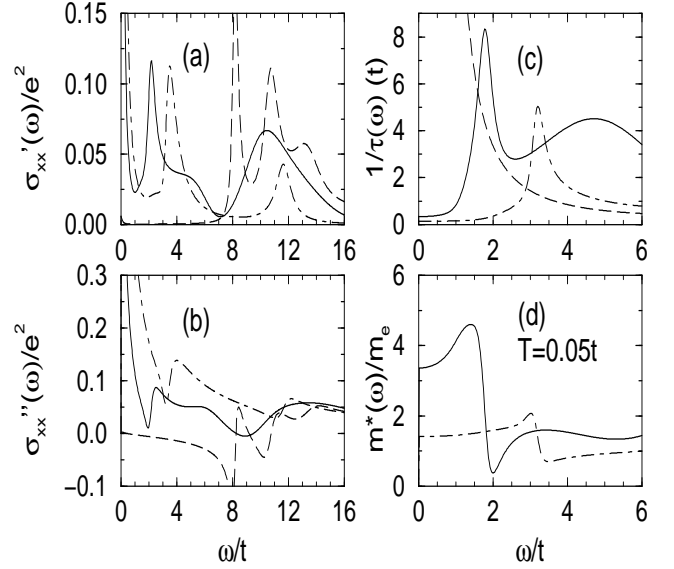,height=3.1in,width=3.3in}}
\narrowtext
\smallskip
\caption
{Optical properties as functions of energy $\omega/t$ for the Hubbard model
 with extended hopping parameters of YBa$_2$Cu$_3$O$_{6+x}$ (Table
 \protect\ref{table1}) at low temperature $T=0.05t$ for
 $\delta=0$    (dashed lines),
 $\delta=0.10$ (  full lines), and
 $\delta=0.25$ (dashed-dotted lines): the meaning of
 different panels is the same as in Fig. \protect\ref{fig:opticlo}. }
\label{fig:opticbi}
\end{figure}

Similar changes in the optical excitation spectra as a function of hole
doping were also found using the effective single-band models with the
parameters representative for La$_{2-x}$Sr$_x$CuO$_4$ (Fig. \ref{fig:opticla})
and YBa$_2$Cu$_3$O$_{6+x}$ (Fig. \ref{fig:opticbi}), respectively. Due to
somewhat larger values of the effective $U$, the gap in the optical spectra
increases to $\sim 6.5t$ and $\sim 7.1t$ in these two compounds. One finds
again that the Drude weight and the mid-gap state appear in the conductivity
of doped systems. For the parameters of La$_{2-x}$Sr$_x$CuO$_4$
(YBa$_2$Cu$_3$O$_{6+x}$) the region of suppressed scattering extends up to
$\simeq 1.3t$ ($\simeq 1.8t$) at $\delta=0.125$. This regime of low $\omega$
gives an enhanced effective mass $\sim 4m_e$ for both sets of model 
parameters. At larger doping $\delta=0.25$ the coherence of the charge 
carriers is enhanced by $t^{\prime}$ and $t^{\prime\prime}$ hopping, and one 
finds a significantly reduced effective scattering between charged carriers, 
extending with roughly no structure over a rather broad energy range. 
Simultaneously, the effective mass $\sim 1.5m_e$ is only little enhanced at 
low energies.

The overall shape of $\sigma^\prime(\omega)$ (Fig. \ref{fig:opticla}) shows
a qualitatively similar behavior to the optical conductivity of
La$_{2-x}$Sr$_x$CuO$_4$ reported by Uchida {\it et al.}.\cite{Uch91} At low
doping the mid-gap band centered at $\omega\simeq 1.7t$ (corresponding to
$0.53$ eV for $J=125$ meV and the present parameters with $J=0.4t$) is
clearly distinguishable from the Drude contribution. It moves to higher
energy $\omega\simeq 2.2t$ (0.7 eV) at $\delta=0.25$. It is quite remarkable 
that our DMFT calculations reproduce qualitatively the structures observed 
in the frequency dependent effective scattering rate $1/\tau(\omega)$ and in 
the effective mass $m^*(\omega)/m_e$ of La$_{2-x}$Sr$_x$CuO$_4$.\cite{Uch91} 
In particular, the strong doping dependence of $1/\tau(\omega)$ and
$m^*(\omega)/m_e$ show the same trends, namely a pronounced reduction of
scattering and effective carrier mass for the heavily doped systems, and 
further justifies the importance of extended hopping parameters in the
cuprates. This behavior originates from an increase of
QP weight in the single-particle excitation spectrum induced by doping.

Puchkov {\it et al.}\cite{Puc96} reported extensive studies of the infrared
properties of YBa$_2$Cu$_3$O$_{6+x}$, Bi$_2$Sr$_2$CaCu$_2$O$_{8+x}$, and
other high-$T_c$ compounds. They found that the far-infrared effective
scattering rate $1/\tau(\omega)$ and the effective mass $m^*(\omega)/m_e$
differ significantly between underdoped and optimally doped samples above
$T_c$. The optimally doped samples show a structureless and lower effective
scattering rate $1/\tau(\omega)$ and a nearly constant and unrenormalized
mass $m^*(\omega)$. On the contrary, in the underdoped samples the
scattering between the charged carriers below $\approx 0.12$ eV is strongly
suppressed and $m^*(\omega)/m_e$ is enhanced in the low-energy region. These
observations are in remarkably good agreement with our findings and supports
our conclusion that the observed doping dependence of $1/\tau(\omega)$ and
$m^*(\omega)/m_e$ originate from an increased coherence of the one-particle
excitation spectra reported in \ref{sec:spectra}, and experimentally observed
in ARPES spectra of Bi$_2$Sr$_2$CaCu$_2$O$_{8+x}$ by Kim {\it et
al.}.\cite{Kim98} The suppression of $1/\tau(\omega)$ below $\omega\simeq
0.12$ eV originates from the opening of a pseudogap in the one-particle
excitation spectrum. Using $J=125$ meV and the value of $J=t/3$ adequate for
YBa$_2$Cu$_3$O$_{6+x}$, we find the energy threshold below which QP
scattering is strongly suppressed in the weakly doped system at $\simeq 0.68$
eV. Unfortunately, this is about a factor of five larger than the
experimental value for underdoped YBa$_2$Cu$_3$O$_{6+x}$ being $\sim J$.
Similar discrepancies in the energy of the QP state with momentum $(\pi,0)$
were reported in Sec. \ref{sec:spectra}. 

\subsection{Drude weight and spectral weight transfer}
\label{sec:drude}

\begin{figure}
\centerline{\psfig{figure=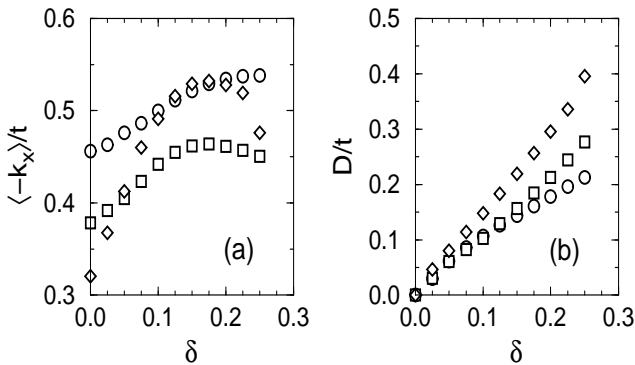,height=2.0in,width=3.3in}}
\narrowtext
\smallskip
\caption
{Kinetic energy along the $x$-direction $\langle -k_x\rangle/t$ and the Drude
 weight $D/t$ as functions of the hole doping $\delta$ for representative
 values of parameters given in Table \protect\ref{table1}:
 Hubbard model (circles),
 La$_{2-x}$Sr$_x$CuO$_4$ (squares), and
 YBa$_2$Cu$_3$O$_{6+x}$ (diamonds).}
\label{fig:drude}
\end{figure}

Finally, we compare the Drude weight $D$ (\ref{Druden}) and the kinetic
energy density associated with $x$-oriented links $\langle -k_x\rangle$
(\ref{xkin}) for the three model parameter sets in Fig. \ref{fig:drude}. The
DMFT gives $\langle -k_x\rangle=0.46t$ for the Hubbard model at half-filling
($\delta=0$) with $U=8t$ which is by a factor $\sim 1.81$ smaller than the HF
result and is in excellent agreement with the value of $\langle -k_x\rangle
=0.49t$ obtained in QMC calculations.\cite{Bul94} We also found an overall
satisfactory agreement of $\langle -k_x\rangle$ as a function of doping with
ED data of Dagotto {\it et al.}.\cite{Dag92} The kinetic energy $\langle
-k_x\rangle$ increases with doping not only because the actual carrier
density changes, but also as a consequence of changing wave vector in the SS
state ${\bf Q}=[\pi(1\pm 2\eta),\pi(1\pm 2\eta)]$ with a gradually increasing 
pitch $\eta$ allowing for coherent electronic transport through the system.
In agreement with QMC data,\cite{Duf95} we observe that
increased extended hopping amplitudes accelerate the SS formation and result
in a stronger increase of $\langle -k_x\rangle$ with $\delta$ for the
YBa$_2$Cu$_3$O$_{6+x}$ than for La$_{2-x}$Sr$_x$CuO$_4$ model parameters.
The $x$-directed kinetic energy shows a linear doping dependence in the
regime of low hole doping, and $\langle -k_x\rangle$ changes by a factor of
$\sim 1.66$ ($1.22$) with respect to half-filling in the case of the
YBa$_2$Cu$_3$O$_{6+x}$ (La$_{2-x}$Sr$_x$CuO$_4$) model parameters, one finds
an faster increase of total spectral weight in the case of stronger hopping
to second and third neighbors, as realized in YBa$_2$Cu$_3$O$_{6+x}$.

The calculated total optical spectral weights are $\propto\langle -k_x\rangle$
following the optical sum rule (\ref{sumo}), and we made a quantitative
comparison with the experimental data. The doping dependence of the total
integrated spectral weight below the charge-transfer band edge at 1.5 eV
reported by Cooper {\it et al.} \cite{Coo90} for La$_{2-x}$Sr$_x$CuO$_4$ is
strikingly similar to the numerical data of Fig. \ref{fig:drude}. The model
reproduces a rapid increase of spectral weight up to $\sim 10\%$ Sr-doping
and a rather doping independent spectral weight in the range of
$0.1\leq x\leq 0.2$. The increase of $\propto\langle -k_x\rangle$ with
increasing doping is faster for the parameters of YBa$_2$Cu$_3$O$_{6+x}$,
with the integrated spectral weight increased by $\sim 1.7$ at $\delta=0.18$
with respect to its value at $\delta=0$.
This value compares again very well, taking the simplicity of the effective
single-band Hubbard model, with a factor of $\sim 1.8$ found by
Orenstein {\it et al.} \cite{Ore90} in the compound with highest $T_c$.

At $\delta=0$ we find a vanishing Drude weight for all three sets of model
parameters, and the system is an insulator. This is of course an expected
result at half-filling, but in the present context it serves as a test of
the internal consistency of theory, as the kinetic energy term
$\langle -k_x\rangle\neq 0$ in Eq. (\ref{Druden}), and has to be compensated
by the current-current correlation function $\Lambda_{xx}({\bf q}=0,2\pi iT)$
in the limit of low temperature. At small hole doping we observed an almost
perfect linear increase of the Drude weight with $\delta$ for all three sets
of model parameters which is an indication of strong electron correlations 
near the Mott insulator at half-filling.\cite{Cas92,Esk94}. 
Such a behavior is compatible with 
a picture of a dilute hole gas in a background with SS LRO which contributes 
to the optical response. However, the crossover to a metal due to increasing 
doping has been analyzed recently using scaling theory,\cite{Ima94} and ED 
technique combined with scaling theory,\cite{Tsu98} which give 
$D\propto\delta^2$ for small doping concentration $\delta$ in a 2D $t$-$J$ 
model. This last result is in sharp contrast to the present picture of a 
dilute hole gas in an AF or SS background, and might indicate that other 
correlations are realized in the spin background when the system is doped, 
namely that the dilute hole gas is unstable towards microscopic phase 
separation, such as realized in polaronic solutions or stripe phases.

The present results demonstrate a substantial transfer of spectral weight to
low energy in the doped systems. We already pointed out earlier \cite{Fle98} 
that the spectral weight transferred into the LHB in the one-particle spectra
agrees with the predictions of perturbation theory in the strongly correlated
regime.\cite{Esk94} In the optical spectra for the Hubbard model at $U/t=8$ 
one finds that the weight transferred into the region below the Mott-Hubbard 
gap is increased by a factor $\sim 1.3$ with respect to $\delta=0.125$ when 
the system is doped to $\delta=0.25$. This change is significant as the total
weight obtained from (\ref{sumo}) via $\langle -k_x\rangle$ remained roughly
constant [see Fig. \ref{fig:drude}(a)], indicating a spectral weight transfer
from the high- to the low-energy region in the single-particle excitation
spectrum.\cite{Esk94} In particular, the weight transfer is in favor of the
Drude weight [Fig. \ref{fig:drude}(b)], which increased in the same doping
range by a factor $\sim 2.15$ although the hole density increased only by a
factor $2$. These changes in the coherent optical weight are consistent with
the observation made in Sec. \ref{sec:spectra} that the single-particle
excitation spectra become {\it more coherent\/} as the hole density increases.

\section{Summary and conclusions}
\label{sec:summa}

We reported a generalization of the DMFT to the magnetically ordered states
and showed that this method allows for a very transparent study of spectral
properties of the Hubbard model at and close to half-filling. The crucial 
step is the derived formula for the self-energy using the Berk-Schrieffer 
\cite{Ber66} spin-fluctuation exchange interaction with an effective 
potential due to particle-particle scattering.\cite{Che91} We have 
demonstrated that this treatment of the many-body effects reproduces the
leading dependence on doping and temperature and gives a very favorable 
comparison with the available numerical data obtained in the QMC and ED 
calculations for a 2D Hubbard model. Although the ${\bf k}$-dependence of the 
self-energy was not included, the spectral functions for a single hole in a 
Mott-Hubbard insulator agrees well with the known structure for the $t$-$J$ 
model,\cite{Dag94} and gives the PES spectrum consisting of a {\it coherent\/} 
QP peak with a dispersion $\sim 2J$, and an {\it incoherent\/} part of width 
$\sim 7t$ at lower energies. We have verified that the QP weight agrees well 
with the ED data in the range of $J/t<0.7$,\cite{Mar91} and supports the 
string picture.\cite{Ede91} Furthermore, the calculation reveals a nontrivial 
relation between the electron occupation factors, $\langle n_{\bf k}\rangle$, 
and the QP weights, $a_{\bf k}$, and shows that the maximum of $a_{\bf k}$ is 
shifted away from the $(\pi/2,\pi/2)$ point, in agreement with the ED results 
of Eskes and Eder.\cite{Esk96}  

Our study has shown that doping of a Mott-Hubbard insulator leads to an
incommensurate magnetic order at low temperatures, which depends on the
actual values of the hopping parameters and the Coulomb interaction $U$.
This kind of magnetic order induces a {\it pseudogap in the one-particle 
spectra\/} which is one of the generic features of the doped Mott-Hubbard 
insulators. The dependence of the pseudogap on the incommensurate magnetic 
order explains why it could not be observed in ED data on small clusters at 
finite temperature,\cite{Jak94} or in the infinite-dimensional Hubbard model 
in the paramagnetic state.\cite{Pru93} This new energy scale due to a {\it 
pseudogap of magnetic origin\/} demonstrates a combination of physics arising 
from the Slater picture and the Mott-Hubbard description of strongly 
correlated electron systems.

The coherent QP states survive in the doped systems, in agreement with the
QMC and ED results. However, the numerical studies suggest that a strong
${\bf k}$-dependence of self-energy might be necessary to describe the 
spectra, as the QP dispersions change. This failure of the rigid QP band
picture finds here quite a different explanation: {\it the changes of the
QP dispersion follow from the incommensurate magnetic order\/} which develops
with doping, and the leading effects in the hole dynamics are still
{\it captured by a local self-energy\/}. 

The one-particle and optical spectra are interrelated, and the opening of a
pseudogap at low temperatures leads to a mid-gap state next to the Drude peak 
in the optical conductivity, both with growing intensity under increasing 
doping. Such features, observed in the SS states at low temperatures, as the 
suppressed scattering rate and large effective mass in the underdoped regime, 
and almost no enhancement of the effective mass in a broad energy range in 
overdoped systems are in remarkably good {\em qualitative\/} agreement with 
the experimental findings in the cuprates.\cite{Uch91,Puc96} This is 
consistent with the reduced density of states $N(\mu)$ at the Fermi energy at 
low temperature. With increasing temperature the value of $N(\mu)$ increases, 
which could not be explained in paramagnetic calculations performed within 
the DMFT approach. It should be realized that such a strong temperature 
dependence of $N(\mu)$ should have important consequences for several 
measurable quantities in the normal phase, as for example Knight shift.

Here we limited ourselves to the qualitative consequences of the extended
hopping $t^{\prime}$ and $t^{\prime\prime}$ for the one-particle and optical
spectra. First of all, the QP dispersion is strongly influenced by these
parameters, and at half filling reproduces the experimental width and 
dispersion of the QP band in Sr$_2$CuO$_2$Cl$_2$.\cite{Wel95} Second, the 
deviation of the characteristic ${\bf Q}$-vector from the AF vector 
$(\pi,\pi)$ increases as a function of doping in SS states, and this process 
is accelerated by a finite value of second- ($t'$)and third-neighbor ($t''$) 
hopping. This explains why the systems with extended hopping are more 
metallic which is indicated by the low effective mass and larger Drude weight.

The dependence of the magnetic order on temperature has also rather drastic
consequences for the measurable quantities. The onset of magnetic order 
below a characteristic temperature results in quite different one-particle
and optical spectra at low temperatures from those obtained in a paramagnetic 
phase. The changes of the spiral ${\bf Q}$-vector with decreasing temperature 
allow to introduce a crossover temperature $T^*$, below which the low-lying 
excitations are gradually modified along with the changes in local magnetic 
order. Such a modification gives a quasi-linear resistivity, and verifies the 
conjecture of Shraiman and Siggia.\cite{Shr89}

In spite of very good agreement for the undoped systems, however, we 
identified several important features which do not agree with the experiments
in the doped cuprates even on a qualitative level, that might indicate that 
either a more accurate treatment of the many-body problem is necessary, or 
more complex magnetic structures are stabilized in these compounds:
  (i) The value of the pseudogap in the one-particle spectra and the 
accompanying energy scale for the suppressed scattering rate in the optical 
conductivity are overestimated by a factor close to five with respect to the 
experimental observations.
 (ii) The SS(1,1) state obtained for the La$_{2-x}$Sr$_x$CuO$_4$ model 
parameters leads to a different splitting of the magnetic scattering peak in 
neutron experiments than the experimentally observed (1,0) and (0,1)
splittings.\cite{Tra97}
(iii) The incommensurate order deviates too fast from the AF state for the
model parameters of YBa$_2$Cu$_3$O$_{6+x}$ which results in different
spin-spin correlations than those observed in experiment, and a QP peak at
the $X$-point moving to too low energies.\cite{Mar96}
 (iv) The doping behavior of the pseudogap and of the related crossover
temperature $T^*$ is opposite to the one observed in the cuprates. In these 
materials the pseudogap and $T^*$ decrease upon doping, whereas here the 
corresponding quantities increase from the $\delta=0.125$ to the $\delta=0.25$ 
case. With increasing doping charge fluctuations become more and more
important which gives rise to the suppression of magnetic order, and 
consequently the pseudogap closes. However, such correlations are 
{\em underestimated\/} in the present treatment, and one finds instead 
a persisting pseudogap.
  (v) Finally, the spiral spin ordering in the (1,1) direction contrasts with 
experimental evidences from neutron scattering in the cuprates, suggesting 
that stripe ordering might play a prominent role in these systems at very low 
temperatures.\cite{Tra97,Yam97} We have found a phase separation at low 
doping levels and therefore the presently studied dilute hole gas in SS 
states is unstable towards magnetic polarons or stripe phases at doping levels
lower than $\delta\simeq 0.1$. 
This motivates a further search for more complex magnetic ground states with 
incommensurate order, and more accurate methods to describe them in theory.

Summarizing, we presented a successful formulation of the DMFT for strongly
correlated magnetic systems, which opens a possibility of further 
applications in transition metal oxides. In contrast to the earlier 
formulations based on the modified second-order formula for the 
self-energy,\cite{Geo96} the present self-energy which describes the 
dynamical effects in the propagation of a hole coupled to spin fluctuations 
allows to obtain stable magnetic solutions: AF ordering at half-filling and 
SS in doped systems. Although it is likely that better variational states, 
possibly with stripe ordering,\cite{Tra97,Yam97,Zaa98} could be found it is 
expected that the presented spectral and optical properties are generic for 
strongly-correlated systems with incommensurate order parameter. A better 
understanding of the cuprates, however, requires a further development of 
theory which should be able to capture the gradual changes of local magnetic 
correlations in doped Mott-Hubbard systems under increasing temperature.

\acknowledgments

It is our pleasure to thank E. Arrigoni, P. Horsch and D. Munzar for
valuable discussions. A.M.O. acknowledges the support by the Committee
of Scientific Research (KBN) of Poland, Project No. 2 P03B 175 14.


\end{multicols} 

\end{document}